\documentclass[12pt]{article}
\usepackage{mathrsfs}
\usepackage{amsmath,amssymb}

\usepackage{amsfonts}
\usepackage{latexsym}
\usepackage{amsthm}
\usepackage{pictex}

\usepackage[cp1251]{inputenc} 
\usepackage{graphicx}

\newcommand{\er}[1]{{\rm(\ref{#1})}}
\def\lb{\label}
\theoremstyle{plain}
\newtheorem{theorem}{\bf Theorem}[section]
\newtheorem{lemma}[theorem]{\bf Lemma}
\theoremstyle{remark}

\newtheorem{corollary}[theorem]{\bf Corollary}
\newtheorem{proposition}[theorem]{\bf Proposition}

\begin{document}

\def\a{\alpha} \def\cA{{\cal A}} \def\bA{{\bf A}}  \def\mA{{\mathscr A}}
\def\b{\beta}  \def\cB{{\cal B}} \def\bB{{\bf B}}  \def\mB{{\mathscr B}}
\def\g{\gamma} \def\cC{{\cal C}} \def\bC{{\bf C}}  \def\mC{{\mathscr C}}
\def\G{\Gamma} \def\cD{{\cal D}} \def\bD{{\bf D}}  \def\mD{{\mathscr D}}
\def\d{\delta} \def\cE{{\cal E}} \def\bE{{\bf E}}  \def\mE{{\mathscr E}}
\def\D{\Delta} \def\cF{{\cal F}} \def\bF{{\bf F}}  \def\mF{{\mathscr F}}
\def\c{\chi}   \def\cG{{\cal G}} \def\bG{{\bf G}}  \def\mG{{\mathscr G}}
\def\z{\zeta}  \def\cH{{\cal H}} \def\bH{{\bf H}}  \def\mH{{\mathscr H}}
\def\e{\eta}   \def\cI{{\cal I}} \def\bI{{\bf I}}  \def\mI{{\mathscr I}}
\def\p{\psi}   \def\cJ{{\cal J}} \def\bJ{{\bf J}}  \def\mJ{{\mathscr J}}
\def\vT{\Theta}\def\cK{{\cal K}} \def\bK{{\bf K}}  \def\mK{{\mathscr K}}
\def\k{\kappa} \def\cL{{\cal L}} \def\bL{{\bf L}}  \def\mL{{\mathscr L}}
\def\l{\lambda}\def\cM{{\cal M}} \def\bM{{\bf M}}  \def\mM{{\mathscr M}}
\def\L{\Lambda}\def\cN{{\cal N}} \def\bN{{\bf N}}  \def\mN{{\mathscr N}}
\def\m{\mu}    \def\cO{{\cal O}} \def\bO{{\bf O}}  \def\mO{{\mathscr O}}
\def\n{\nu}    \def\cP{{\cal P}} \def\bP{{\bf P}}  \def\mP{{\mathscr P}}
\def\r{\rho}   \def\cQ{{\cal Q}} \def\bQ{{\bf Q}}  \def\mQ{{\mathscr Q}}
\def\s{\sigma} \def\cR{{\cal R}} \def\bR{{\bf R}}  \def\mR{{\mathscr R}}
\def\S{\Sigma} \def\cS{{\cal S}} \def\bS{{\bf S}}  \def\mS{{\mathscr S}}
\def\t{\tau}   \def\cT{{\cal T}} \def\bT{{\bf T}}  \def\mT{{\mathscr T}}
\def\f{\phi}   \def\cU{{\cal U}} \def\bU{{\bf U}}  \def\mU{{\mathscr U}}
\def\F{\Phi}   \def\cV{{\cal V}} \def\bV{{\bf V}}  \def\mV{{\mathscr V}}
\def\P{\Psi}   \def\cW{{\cal W}} \def\bW{{\bf W}}  \def\mW{{\mathscr W}}
\def\o{\omega} \def\cX{{\cal X}} \def\bX{{\bf X}}  \def\mX{{\mathscr X}}
\def\x{\xi}    \def\cY{{\cal Y}} \def\bY{{\bf Y}}  \def\mY{{\mathscr Y}}
\def\X{\Xi}    \def\cZ{{\cal Z}} \def\bZ{{\bf Z}}  \def\mZ{{\mathscr Z}}
\def\O{\Omega}
\def\ve{\varepsilon}
\def\vt{\vartheta}
\def\vp{\varphi}
\def\vk{\varkappa}

\newcommand{\gA}{\mathfrak{A}}
\newcommand{\gB}{\mathfrak{B}}
\newcommand{\gC}{\mathfrak{C}}
\newcommand{\gD}{\mathfrak{D}}
\newcommand{\gE}{\mathfrak{E}}
\newcommand{\gF}{\mathfrak{F}}
\newcommand{\gG}{\mathfrak{G}}
\newcommand{\gH}{\mathfrak{H}}
\newcommand{\gI}{\mathfrak{I}}
\newcommand{\gJ}{\mathfrak{J}}
\newcommand{\gK}{\mathfrak{K}}
\newcommand{\gL}{\mathfrak{L}}
\newcommand{\gM}{\mathfrak{M}}
\newcommand{\gN}{\mathfrak{N}}
\newcommand{\gO}{\mathfrak{O}}
\newcommand{\gP}{\mathfrak{P}}
\newcommand{\gR}{\mathfrak{R}}
\newcommand{\gS}{\mathfrak{S}}
\newcommand{\gT}{\mathfrak{T}}
\newcommand{\gU}{\mathfrak{U}}
\newcommand{\gV}{\mathfrak{V}}
\newcommand{\gW}{\mathfrak{W}}
\newcommand{\gX}{\mathfrak{X}}
\newcommand{\gY}{\mathfrak{Y}}
\newcommand{\gZ}{\mathfrak{Z}}

\def\mA{{\mathscr A}}
\def\mB{{\mathscr B}}
\def\mC{{\mathscr C}}
\def\mD{{\mathscr D}}
\def\mE{{\mathscr E}}
\def\mF{{\mathscr F}}
\def\mG{{\mathscr G}}
\def\mH{{\mathscr H}}
\def\mI{{\mathscr I}}
\def\mJ{{\mathscr J}}
\def\mK{{\mathscr K}}
\def\mL{{\mathscr L}}
\def\mM{{\mathscr M}}
\def\mN{{\mathscr N}}
\def\mO{{\mathscr O}}
\def\mP{{\mathscr P}}
\def\mQ{{\mathscr Q}}
\def\mR{{\mathscr R}}
\def\mS{{\mathscr S}}
\def\mT{{\mathscr T}}
\def\mU{{\mathscr U}}
\def\mV{{\mathscr V}}
\def\mW{{\mathscr W}}
\def\mX{{\mathscr X}}
\def\mY{{\mathscr Y}}
\def\mZ{{\mathscr Z}}

\def\Z{{\Bbb Z}}
\def\R{{\Bbb R}}
\def\C{{\Bbb C}}
\def\T{{\Bbb T}}
\def\N{{\Bbb N}}
\def\S{{\Bbb S}}
\def\H{{\Bbb H}}
\def\J{{\Bbb J}}

\def\qqq{\qquad}
\def\qq{\quad}
\newcommand{\ma}{\begin{pmatrix}}
\newcommand{\am}{\end{pmatrix}}
\newcommand{\ca}{\begin{cases}}
\newcommand{\ac}{\end{cases}}
\let\ge\geqslant
\let\le\leqslant
\let\geq\geqslant
\let\leq\leqslant
\def\ma{\left(\begin{array}{cc}}
\def\am{\end{array}\right)}
\def\iint{\int\!\!\!\int}
\def\lt{\biggl}
\def\rt{\biggr}
\let\geq\geqslant
\let\leq\leqslant
\def\[{\begin{equation}}
\def\]{\end{equation}}
\def\wh{\widehat}
\def\wt{\widetilde}
\def\pa{\partial}
\def\sm{\setminus}
\def\es{\emptyset}
\def\no{\noindent}
\def\ol{\overline}
\def\iy{\infty}
\def\ev{\equiv}
\def\/{\over}
\def\ts{\times}
\def\os{\oplus}
\def\ss{\subset}
\def\h{\hat}
\def\Re{\mathop{\rm Re}\nolimits}
\def\Im{\mathop{\rm Im}\nolimits}
\def\supp{\mathop{\rm supp}\nolimits}
\def\sign{\mathop{\rm sign}\nolimits}
\def\Ran{\mathop{\rm Ran}\nolimits}
\def\Ker{\mathop{\rm Ker}\nolimits}
\def\Tr{\mathop{\rm Tr}\nolimits}
\def\const{\mathop{\rm const}\nolimits}
\def\dist{\mathop{\rm dist}\nolimits}
\def\diag{\mathop{\rm diag}\nolimits}
\def\Wr{\mathop{\rm Wr}\nolimits}
\def\BBox{\hspace{1mm}\vrule height6pt width5.5pt depth0pt \hspace{6pt}}

\def\Diag{\mathop{\rm Diag}\nolimits}


\def\Twelve{
\font\Tenmsa=msam10 scaled 1200 \font\Sevenmsa=msam7 scaled 1200
\font\Fivemsa=msam5 scaled 1200 \textfont\msbfam=\Tenmsb
\scriptfont\msbfam=\Sevenmsb \scriptscriptfont\msbfam=\Fivemsb

\font\Teneufm=eufm10 scaled 1200 \font\Seveneufm=eufm7 scaled 1200
\font\Fiveeufm=eufm5 scaled 1200
\textfont\eufmfam=\Teneufm \scriptfont\eufmfam=\Seveneufm
\scriptscriptfont\eufmfam=\Fiveeufm}

\def\Ten{
\textfont\msafam=\tenmsa \scriptfont\msafam=\sevenmsa
\scriptscriptfont\msafam=\fivemsa

\textfont\msbfam=\tenmsb \scriptfont\msbfam=\sevenmsb
\scriptscriptfont\msbfam=\fivemsb

\textfont\eufmfam=\teneufm \scriptfont\eufmfam=\seveneufm
\scriptscriptfont\eufmfam=\fiveeufm}

\title {Zigzag and armchair nanotubes in external fields}

\author{
Evgeny Korotyaev
\begin{footnote}
{School of Mathematics, Cardiff Univ., Senghennydd Road, Cardiff,
CF24 4AG, UK, e-mail: korotyaeve@cf.ac.uk}
\end{footnote}
 \and Anton Kutsenko
\begin{footnote}
{ Department of
 Mathematics of Sankt-Petersburg State University, Russia e-mail: kucenkoa@rambler.u}
\end{footnote}
}

\maketitle

\begin{abstract}
\no We consider the Schr\"odinger operator on the zigzag and
armchair nanotubes
 (tight-binding models) in a uniform magnetic field
 $\mB$ and  in an external periodic electric potential.
 The magnetic and electric fields are parallel to the axis of the  nanotube.
We show that this operator is unitarily equivalent to the finite
orthogonal sum of Jacobi operators. We describe all spectral bands
and all eigenvalues (with infinite multiplicity, i.e., flat bands).
 Moreover, we determine the asymptotics of the spectral bands both
for small and large potentials. We describe the spectrum as a
function of $|\mB|$. For example, if $|\mB|\to {16\/3}({\pi\/2}-{\pi
k\/N}+\pi s)\tan {\pi\/2N}, k=1,2,..,N, s\in \Z$, then some spectral
band for zigzag nanotube shrinks into a flat band and the
corresponding asymptotics are determined.
\end{abstract}


\section {Introduction.}
\setcounter{equation}{0}

After their discovery \cite{Ii}, carbon nanotubes remain in both
theoretical and applied research \cite{SDD}.
Structure of nanotubes are formed by rolling up a graphene sheet
into a cylinder.
Such nanomodels were introduced by Pauling \cite{Pa} in 1936 to
simulate aromatic molecules. They were described in more detail by
Ruedenberg and Scherr \cite{RS1} in 1953.
Various physical properties of carbon nanotubes can be found in
\cite{SDD}.


There are mathematical results about Schr\"odinger operators on
carbon nanotubes (zigzag, armchair and chiral) (see \cite{BK},
\cite{KL}, \cite{KL1}, \cite{K1}, \cite{KuP}, \cite{Pk}). All these
papers consider the so called  continuous  models. But in the
physical literature the most commonly used model is the
tight-binding model.("In solid state physics, the tight binding
model is an approach to the electronic band structure from the
atomic limit case. In the tight binding model, it is assumed that
the Fourier transform of the Bloch function can be approximated by
the Linear Combination of Atomic Orbital(LCAO). Starting from the
Hamiltonian of an isolated atom centered at each lattice point, the
band structure of solids can be investigated.") For applications of
our models see ref. in \cite{ARZ}, \cite{SDD}, \cite{Ha}.

In this paper we concentrate on carbon nanotubes which arise from graphene:
 zigzag and armchair nanotubes (see physical propereties in \cite{SDD}).
We will study and compare spectral properties of Shr\"odinger
operator on zigzag and armchair nanotubes. We will
show that these operators have different spectral properties.

For example:

1) The Shr\"odinger operator $H_{zi}$ on the zigzag nanotube is unitarily
equivalent to the direct sum of scalar Jacobi matrices (see Theorem
\ref{T1}). But the Shr\"odinger operator on armchair nanotube
$H_{ar}$ is unitarily equivalent to the direct sum of Jacobi matrices
with $2\ts 2$ matrix valued coefficients (see Theorem \ref{T1arm}). Then the spectral analysis of $H_{ar}$ is
more difficult.

2) For some amplitude of the constant  magnetic field the spectrum of
$H_{zi}$ has absolutely continuous part and eigenvalues (flat bands,
see Theorem \ref{T2}). But the spectrum of $H_{ar}$ is purely
absolutely continuous for any amplitude of the magnetic field.

3) The spectral bands of operators $H_{zi}$ and $H_{ar}$ are different.
But in some cases the spectra of these operators has the same part
(see Theorem \ref{Tcomp}).

4) In the simple case, when the magnetic field is absent and
external electric potential has minimal period $2$ the spectrum of $H_{zi}$ and $H_{ar}$
are coincide.
Remark that the multiplicity of some spectral zones is different (see Sect 4
and Sect. 6.2).

5) The structure of spectral zones of $H_{ar}$ and $H_{zi}$ for large
electric potentials is similar, since the spectrum is a union of small
clusters, but asymptotics of this clusters are different (see
Theorem \ref{T44} and Theorem \ref{Larmp}). Moreover, we have similar
situation for small potentials.

In the proof of our theorems we determine various asymptotics for
periodic Jacobi operators with specific coefficients see \er{Jk}.
Note that there exist a lot of papers devoted to asymptotics and
estimates both for periodic Jacobi operators and Schr\"odinger
operators see e.g. \cite{KKu1}, \cite{La}, \cite{vMou}, \cite{S1}, \cite{S2}.

\section {Zigzag nanotube.}
\setcounter{equation}{0}

In this Section we consider the
Schr\"odinger operator $H^b$  on the zigzag nanotube $\G\ss\R^3$ (1D
models tight-binding model of zigzag single-wall nanotubes, see
\cite{SDD}, \cite{N}) in a uniform magnetic field $\mB=|\mB|{\bf
e}_0,\ {\bf e}_0=(0,0,1)\in \R^3$ and in an external electric
potential. Our model nanotube $\G$ is a graph (see Fig. \ref{f001}
and 2) embedded in $\R^3$ oriented in the $z$-direction ${\bf e}_0$
with unit edge length. $\G$ is a set of vertices (atoms) ${\bf
r}_{\o}$ connecting by bonds (edges) $\G_{n,k,j}$ and
\begin{multline}
\G=\cup_{\o\in \cZ} {\bf r}_{\o},\qq {\bf r}_{n,0,k}={\bf
\vk}_{n+2k}+{3n\/2}{\bf e}_0,\ \qqq {\bf r}_{n,1,k}={\bf r}_{n,0,k}+{\bf e}_0,\qqq \o=(n,j,k)\in \cZ,\\
\cZ=\Z\ts \{0,1\}\ts \Z_N, \qq \Z_N=\Z/(N\Z),
 \qq {\bf \vk}_k=R(\cos{\pi
k\/N},\sin{\pi k\/N},0),\qq \ R={\sqrt 3\/4\sin {\pi\/2N}}.
\end{multline}
\begin{figure}[h]
         \centering\includegraphics[clip]{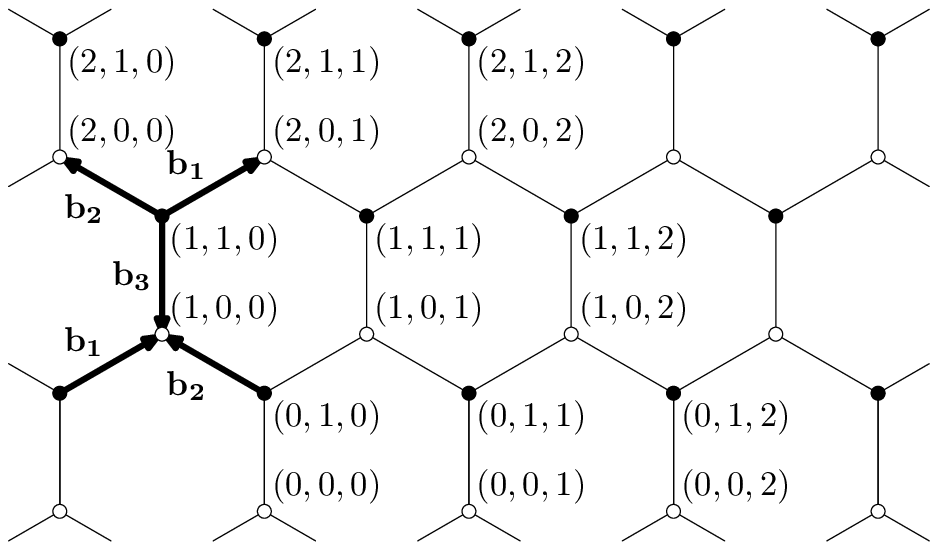}
         \begin{center}
         {\small Fig 1. A piece of zigzag nanotube. }
         \end{center}
         \lb{f001}
\end{figure}
Our carbon model nanotube is the honeycomb lattice of a graphene
sheet rolled into a cylinder. This nanotube $\G$ has $N$ hexagons
around the cylinder embedded in $\R^3$. Here $n \in \Z$ labels the
position in the axial direction of the tube, $j=0,1$ is a label for
the two types of vertices (atoms)  (see Fig. \ref{f001}), and  $k
\in \Z_N$ labels the position around the cylinder. The points ${\bf
r}_{0,1,k}, k\in \Z_N$ are vertices of the regular N-gon $\mP_0$ and
${\bf r}_{1,0,k}$ are the vertices of the regular N-gon $\mP_1$.
$\mP_1$ arises from $\mP_0$ by combination of the rotation around
the axis of the cylinder $\cC$ by the angle ${\pi\/N}$ and of the
translation by  ${1\/2}{\bf e}_0$. Repeating this procedure we
obtain $\G$.

Introduce the Hilbert space $\ell^2(\G)$ of functions
$f=(f_\o)_{\o\in \cZ}$ on $\G$ equipped with the norm
$\|f\|_{\ell^2(\G)}^2=\sum_{\o\in \cZ} |f_\o|^2 $. The tight-binding
Hamiltonian $H^b$ on the nanotube $\G$ has the form $H^b=H_0^b+V$ on
$\ell^2(\G)$, where $H_0^b$ is the Hamiltonian of the nanotube in
the magnetic field and is given by
\begin{multline}
 \lb{010}
 (H_0^b f)_{n,0,k}=e^{ib_2}f_{n-1,1,k}+e^{ib_1}f_{n-1,1,k-1}+
 e^{ib_3}f_{n,1,k},\\
 (H_0^b f)_{n,1,k}=e^{-ib_1}f_{n+1,0,k+1}+e^{-ib_2}f_{n+1,0,k}+
 e^{-ib_3}f_{n,0,k},\qq f=(f_\o)_{\o\in \cZ},\\
 \o=(n,j,k)\in \Z\ts \{0,1\}\ts \Z_{N},\qq
 b_3=0,\ \ b_1=-b_2=b={3|\mB|\/16}\cot {\pi\/2N},
\end{multline}
(the last line in \er{010} was obtained in \cite{KL1}) and the
operator $V$ corresponding to the  external electric potential
 is given by
\[
\lb{cb} (Vf)_\o=V_\o f_\o,\qq where \qq  V_{n-1,1,k}=v_{2n}, \qq
V_{n,0,k}=v_{2n+1},\qq v=(v_n)_{n\in\Z}\in\ell^\iy.
\]
Such potentials can be realized using optical methods, by gating, or
by an acoustic field (see \cite{N}). For example, if an external
potential is given by $A_0\cos (\x_0 z+\b_0)$ for some constant
$A_0,\x_0, \b_0$, then we obtain
\[
v_{2n}=A\cos (2\pi\x (n-{1\/3})+\b),\ \qqq v_{2n+1}=A\cos (2\pi\x
n+\b), \qq n\in \Z,
\]
 for some constant $A,\x, \b$. If $\x$ is rational, then the sequence $v_n,
n\in \Z$ is periodic. If $\x$ is irrational, then the sequence $v_n,
n\in \Z$ is almost periodic.


\begin{figure}[h]
\lb{fig2}
         \centering\includegraphics[clip]{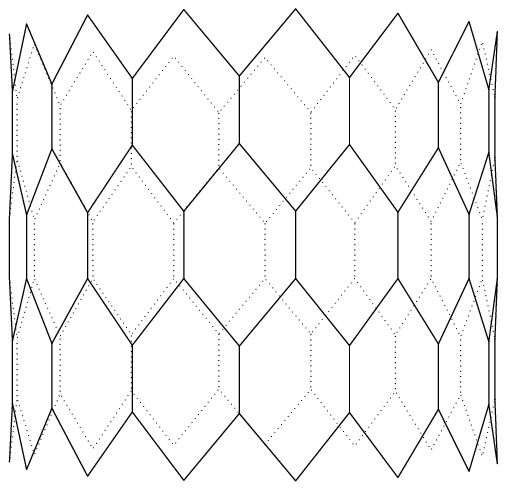}
         \begin{center}
         {\small Fig 2. Nanotube in the magnetic field. }
         \end{center}
\end{figure}

Below we use notation $\N_j$ for the set $\{1,..,j\}$, $j\ge1$.

\begin{theorem}
\label{T1} i) Let $v=(v_n)_{n\in \Z}\in \ell^\iy$. Then each
operator $H^b,b\in \R$ is unitarily equivalent to the operator
$\os_1^N J_k^b$, where $J_k^b$ is a Jacobi operator, acting on
$\ell^2(\Z)$ and given by
\begin{multline}
 \lb{Jk}
 (J_k^b y)_n=a_{n-1}y_{n-1}+a_{n}y_{n+1}+v_ny_n,\qq
 y=(y_n)_{n\in \Z}\in \ell^2,\\
  a_{2n}\ev a_{k,2n}=2|c_k|,\qq  \qq a_{2n+1}\ev a_{k,2n+1}=1,\ \ c_k=\cos (b+{\pi k\/N}), \
 n\in\Z,
\end{multline}
and $J_k^{b+{\pi\/N}}=J_{k+1}^b$, $J_k^{-b}=J_{N-k}^b$ for all
$(k,b)\in\Z_N\ts \R$. Moreover, the operators $H^{\pm b}$ and
$H^{b+{\pi\/N}}$ are unitarily equivalent for all $b\in \R$.

\no  ii) Let, in addition,  $c_k=\cos (b+{\pi k\/N})=0$ for some
$(k,b)\in \Z_N\ts \R$. Then
\[
\lb{spp}
\s(J_k^b)=\s_{pp}(J_k^b)=\rt\{z_{n,j}=v_n^++(-1)^j|{v_n^-}^2+1|^{1\/2},\qq
v_n^\pm={v_{2n-1}\pm v_{2n}\/2} ,\qq (n,j)\in \Z\ts \N_2\rt\}.
\]
\end{theorem}

{\no\bf Remark.} 1)  The matrix of the operator $J_k^b$ is given by
\[
\lb{Jk1a}
 J_k^b=\left(\begin{array}{ccccccc}
 ... &...&... & ...      &...       &...       &... \\
 ... &2|c_k|  &v_1   & 1     &0         & 0        &... \\
 ... &0  & 1& v_2        &2|c_k|& 0        &... \\
 ... &0  & 0  &2|c_k|& v_3         & 1     &... \\
 ... &0  & 0  & 0        &1      & v_4       &... \\
 ... &0  & 0  & 0        &0         &2|c_k|&... \\
 ... &.  & ...& ...      &...       &...       &... \\
 \end{array}\right).
\]
2) If $|c_k|={1\/2}$, then $J_k^b$ is the Schr\"odinger operator
with $a_{n}=1$ for all $n\in \Z$. In particular, if $b=0,{N\/3}\in
\N$, then $J_{N\/3}^0$ is the Schr\"odinger operator.

\no 3)  In the continuous models similar results were obtained in
 \cite{KL}, \cite{KL1}.

\no  4)  Exner \cite{Ex} obtained a duality between Schr\"odinger
 operators on graphs and certain Jacobi matrices, which  depend on energy.
 In our case the   Jacobi matrices do not depend on energy.


\no {\bf 1. Periodic electric potentials $v$}. Introduce the class
$\ell_s^{per}$ of real $s$-periodic sequences $v=(v_n)_{n\in \Z}\in
\ell^\iy$ and $v_{n+s}=v_n$, for all $n\in\Z$.
If $v\in \ell_{p_*}^{per}, p_*\ge 1$, then $J_k^b$ is 2p-periodic
matrix where
\[
\lb{dep}
p=\ca {p_*\/2} & if \ p_*\ \ is \ even\\
p_* & if \ p_* \ \ is \ odd       \ac.
\]
If $c_k\ne 0$ for some $(k,b)\in \Z_N\ts \R$, then  the spectrum of
$J_k^b$ has the form
\begin{multline}
\lb{sJk} \s(J_k^b)=\s_{ac}(J_k^b)=\cup_1^{2p} \s_{k,n}^b,\qq
\s^{b}_{k,n}=[z_{k,n-1}^{b,+},z_{k,n}^{b,-}],\qq  n\in\N_{2p},\\
z_{k,0}^{b,+}<z_{k,1}^{b,-}\le z_{k,1}^{b,+}<z_{k,2}^{b,-}\le
z_{k,2}^{b,+}<...<z_{k,2p}^{b,-},
\end{multline}
see \cite{vM}, where  $z_{k,n}^{b,\pm}$ are 4p-periodic eigenvalues
for the equation $a_{n-1}y_{n-1}+a_{n}y_{n+1}+v_ny_n=z y_n,\
y=(y_n)_{n\in \Z}$. The intervals $\s_{k,n}^b,\s_{k,n+1}^b$ are
separated by  a gap $\g_{k,n}^b=(z_{k,n}^{b,-},z_{k,n}^{b,+})$ of
length $|\g_{k,n}^b|\ge 0$. If a gap $\g_{k,n}^b$ is degenerate,
i.e., $|\g_{k,n}^b|=0$, then the corresponding segments
$\s_{k,n}^b$, $\s_{k,n+1}^b$ merge.

If $c_k=0$ for some $(k,b)\in \Z_N\ts \R$, then \er{spp} gives
$\s(J_k^b)=\s_{pp}(J_k^b)$, where
\[
 \s_{pp}(J_k^b)=\rt\{z_{n,j}=v_n^++(-1)^j\sqrt{{v_n^-}^2+1},\qq
 v_n^\pm={v_{2n-1}\pm v_{2n}\/2} ,\qq (n,j)\in \N_p\ts \N_2\rt\},
\]
and each eigenvalue of $J_k^b$  is  a flat band, i.e. it  has
infinite multiplicity. In Theorem \ref{T2} we show that the spectral
band $\s_{k,n}^b$ shrinks to the flat band $\{\l_n\}$ as $c_k\to 0$
and the corresponding asymptotics are determined.

Each operator $J_k^b$ is unitarily equivalent to the operator
$\int_{[0,2\pi)}^{\os}K(e^{it},a){dt\/2\pi }, a=2|c_k|$, where
$2p\ts 2p$ matrix $K(\t,a)\ev K(\t,a,v)$  is a Jacobi operator,
acting on $\C^{2p}$ and given by
\[
\lb{dK}
 K(\t,a)=K^0(\t,a)+B,\qq
 K^0(\t,a)=\left(\begin{array}{ccccc} 0 & 1 & 0 & ... & {a\/\t} \\
                                   1 & 0 & a & ... & 0 \\
                                  0 &  a & 0 & ... & 0 \\
                                  ... & ... & ... & ... & ... \\
                                  \t a & 0 & ... & 1 & 0
                                  \end{array}\right),\
 B=\diag (v_n)_1^{2p},
\]
where $\t\in \S^1=\{\t\in\C:|\t|=1\}$. Let $\m_{1}(\t,a)\le
\m_{2}(\t,a)\le \m_{3}(\t,a)\le ....\le \m_{2p}(\t,a)$ be
eigenvalues of $K(\t,a),  \t\in \S^1$, here $\m_{n}(\cdot,a)$ is
analytic function in $\t\in \S^1$. Note that
$\m_{n}(\S^1,a)=\s_{k,n}^b$ for all $(k,n)\in \Z_N\ts\N_{2p}$. If
$c_k\ne 0$, then each $\m_{n}(\cdot,a), n\in \N_{2p}$ is not a
constant and $|\s_{k,n}^b|>0$. If $c_k=0$ for some $k\in \Z_N$, then
each $\m_{n}(\cdot,0)=\const=\l_n, n\in \N_{2p}$ and
$\s_{k,n}^b=\{\l_n\}$ is a flat band.

\no {\bf 2. The case $\mB=0$}. Consider the Schr\"odinger operator
$H^0$ at $\mB=0$. By Theorem \ref{T1}, the operator $H^0$ is
unitarily equivalent to the operator $\os_1^N J_k^0$, where $J_k^0$
is a Jacobi operator $J_k^b$ at $b=0$ and here $a_{2n}=2|\cos {\pi
k\/N}|,\ \  a_{2n+1}=1$. Note that if $k\ne {N\/2}$, then
$\s(J_k^0)=\s_{ac}(J_k^0)$
  and if $k={N\/2}$, then  $\s(J_k^0)=\s_{pp}(J_k^0)$.

\no {\bf 3. Example of simple  periodic potentials $v$}. Consider
the potential $v=v_{2k+1}=-v_{2k}\in \R, k\in \Z$. In Section 4 we
will show that
$$
\s(J_k^b)=[z_{k,0}^{b,+},z_{k,0}^{b,-}]\sm \g_{k,1}^b,\qqq
\g_{k,1}^b= (z_{k,1}^{b,-},z_{k,1}^{b,+}),
$$
$$
z_{k,0}^{b,\mp}=\pm \sqrt{v^2+(2|c_k|+1)^2},\qqq z_{k,1}^{b,\pm}=\pm
\sqrt{v^2+(2|c_k|-1)^2},\qq k\in\Z_N,
$$
where $\g_{k,1}^b$ is the gap in the spectrum of $J_k^b$. This gives
$$
\s(J_k^b)=\s_{ac}(J_k^b)\cup \s_{pp}(J_k^b), \qq
\s_{pp}(J_k^b)=\ca\es    & if \  c_k\ne 0,  \\
    \{\pm \sqrt{1+v^2}\}  & if \ c_k=0, \   \ac,
$$
and then we deduce that the spectrum of $H^b$ is given by
\[
\s(H^b)=\s_{ac}(H^b)\cup \s_{pp}(H^b), \qq
\s_{pp}(H^b)=\ca\es    & if \  c_k\ne 0, \ any \ k\in \Z_N \\
    \{\pm \sqrt{1+v^2}\}  & if \ c_k=0, \ some \ k\in \Z_N
    \ac,
\]
\[
\s_{ac}(H^b)=[z_{0}^{b,+},z_{0}^{b,-}]\sm \g(H^b),\ \g(H^b)=
(z_{1}^{b,-},z_{1}^{b,+}),
\]
where $\g(H^b)$ is the gap in the spectrum of $H^b$. Note that if
$c_k=0$ for some $k\in\Z_N$ then $\s_{pp}(H^b)=\{\pm
\sqrt{1+v^2}\}\ss \g(H^b)$. Theorem \ref{T1}.i yields
$\s(H^{b+{\pi\/N}})=\s(H^b )$ for all $b\in \R$. Then we need to
consider only the case $b\in [0,{\pi\/N})$ and in this case we get
\[
z_{0}^{b,+}=\ca z_{0,0}^{b,+}    & if  \qq b\le{\pi\/2N}\\
    z_{N-1,0}^{b,+}  & if  \qq b>{\pi\/2N}  \ac.
\]
Moreover, in particular case $\mB=0$ we obtain
\[
\g(H^0)=(-|v|, |v|),\qq \qq  if \qqq  {N\/3}\in \N,\  b=0.
\]
Now we return to the general case of periodic potentials. First
theorem is devoted to the asymptotics of small spectral bands that
degenerate to the flat band.

\begin{theorem}
\lb{T2} Let $v\in \ell_{p_*}^{per}$, $p_*\ge1$ and $c_k\to 0$ as
$b\to b_0={\pi\/2}-{\pi k\/N}$ for some $k\in \Z_N$ and let
$\l_1\le\l_2\le..\le\l_{2p}$ be eigenvalues of $K(1,0,v)$. Then the
endpoints $z_{k,s-1}^{b,+},z_{k,s}^{b,-}$, $s\in\N_{2p}$ of the
spectral bands $\s_{k,s}^{b}=[z_{k,s-1}^{b,+},z_{k,s}^{b,-}]$ are
analytic functions in $b\in \{|b-b_0|<\ve\}$ for some $\ve>0$ and
satisfy
\[
\lb{T2-1}
 z_{k,s-1}^{b,+}=\l_s+O(c_k^2),\ \ z_{k,s}^{b,-}=\l_s+O(c_k^2)
 \qq as \qq c_k\to 0.
\]
Let in addition $\l_{s-1}<\l_s<\l_{s+1}$ for some $s\in\N_{2p}$,
where $\l_0=-\iy$, $\l_{2p+1}=+\iy$. Then
\[
\lb{T2-2} z_{k,s}^{b,-}=\l_s- {2\/\L_s}|2c_k|^p+ \sum_{2\le 2n\le
p}C_{k,n}(2c_k)^{2n} +O(c_k^{p+1}),\qq \L_s=\prod_{n=1,\
 n\not=s}^{2p}|\l_s-\l_n|,
\]
\[
\lb{T2-3}
 |\s_{k,s}^b|=z_{k,s}^{b,-}-z_{k,s-1}^{b,+}={4|2c_k|^p\/\L_s}+O(c_k^{p+1})
\]
as $c_k\to 0$ for some constants $C_{k,n}$, which depend only  on
$v$.
\end{theorem}

\no {\bf Remark.} By \er{T2-1}, each spectral band $\s_{k,n}^b, n\in
\N_{2p}$ shrinks to the flat band $\{\l_n\}$ as $c_k\to 0$.

We consider the nanotube in {\bf weak electric fields}. Our operator
has the form  $H^b(t)=H_0^b+t V$, where a coupling constant $
t\to0$. In this case the corresponding Jacobi operator depend on $t$
and is given by
\[
\lb{j000} (J_k^b(t)y)_n=a_{n-1}y_{n-1}+a_{n}y_{n+1}+tv_ny_n,\
y=(y_n)_{n\in \Z}\in \ell^2,    \ n\in\Z,
\]
and $a_{2n}=2|c_k|,\ a_{2n+1}=1$. We study how the spectral bands
$\s_{k,n}^b(t)=[z_{k,n-1}^{b,+}(t),z_{k,n}^{b,-}(t)]$, $n\in\N_{2p}$
of the operator $J_k^b(t) $   depend on the couple constant $t\to
0$.

For $v\in \ell_{p_*}^{per}$ we define two vectors
$v^0=(v_{2n})_{1}^{p}, \ v^1=(v_{2n-1})_{1}^{p}\in \R^p$ and
\begin{multline}
 \lb{j010}
 \hat u_n=\langle
u,e_n\rangle,\ \ u\in \C^p,\ \ e_n={1\/2p}(\t_n^{2j})_{j=1}^p\in
\C^p, \ \ \t_n=e^{i\frac{\pi n}p},\ \  \hat u_{p+n}=\hat u_{p-n},\ \
n\in\N_{p}.
\end{multline}
Here $e_n, n\in \N_p $ is a basis in $\C^p$ and $\langle
u,e_n\rangle$ is the scalar product in $\C^p$. Define
$\ell_{0,p_*}^{per}=\{v\in \ell_{p_*}^{per}: \sum_1^{2p}v_n=0\}$ and
the sets
$\N_{k,p}=\ca\N_{2p-1}& if\  2|c_k|=1\\
\N_{2p-1}\sm\{p\} & if\  2|c_k|\ne1\ac $.

\begin{theorem}
\lb{T3} Let $c_k\ne 0$ for some $(k,b)\in \Z_N\ts \R$. Let $v\in
\ell_{0,p_*}^{per}$ and let $v^0=(v_{2n})_{1}^{p}, \
v^1=(v_{2n-1})_{1}^{p}\in \R^p$. Then the asymptotic of the spectral
bands $\s_{k,n}^b(t)=[z_{k,n-1}^{b,+}(t),z_{k,n}^{b,-}(t)],
n\in\N_{2p}$ of the operator $J^{b}_k(t)$ hold true
\begin{multline}
\lb{T3-1}
 z_{k,n}^{b,\pm}(t)=z^{\pm}_{n,k}(0)\pm t\p_{k,n}(v)+O(t^2),\ \
 n\in \N_{k,p},\\
 \p_{k,n}(v)=\ca|\hat v_n^0+e^{2i\arg(2|c_k|+\t_n)}\hat v_n^1|,\ \ n\ne p\\
                |\hat v_p^0-\hat v_p^1|,\ \ 2|c_k|=1,\ \ n=p\ac,
\end{multline}
\begin{multline}
\lb{T3-2} z_{k,0}^{b,+}(t)=z_{k,0}^{b,+}(0)+O(t^2),\ \
 z_{k,2p}^{b,-}(t)=z_{k,2p}^{b,-}(0)+O(t^2),\\
and \qq \ if\ 2|c_k|\ne1 \qqq \Rightarrow \qq
z_{k,p}^{b,\pm}(t)=z_{k,p}^{b,\pm}(0)+O(t^2),
\end{multline}
\[
\lb{T3-3}
 z_{k,n}^{b,\pm}(0)=|2|c_k|+\t_n|\sign(n-p),\ \
 n\in\N_{2p-1}\sm\{p\},\ \
 z_{k,p}^{b,\pm}(0)=\pm|2|c_k|-1|,
\]
as $t\downarrow 0$. Moreover, if $p_*$ is odd, then for all $n\in
\N_{k,p}$ the following identities hold true
\begin{multline}
 \lb{T3-5a}
\hat v_n^0=\t_n^{p+1}\hat v_n^1,\qqq
 \p_{k,n}(v)=|\hat v_n^0|\r_{k,n},\ \
\\
 \r_{k,n}=\ca |(-1)^n\t_n+e^{2i\arg(2|c_k|+\t_n)}|,\ n\ne p\\
                   0,\ if\ 2|c_k|=1\ and\ n=p\ac,
 \ca \r_{k,n}\ne0,\ \ \ if\ |c_k|\ne\frac12,\ \ \\
     \r_{k,n}\ne0,\ \ \ if\ |c_k|=\frac12,\ \ even\ \ n\\
     \r_{k,n}=0,\ \  \ if\ |c_k|=\frac12,\ \ odd\ \ n\ac.
\end{multline}
\end{theorem}

\no To describe some examples of external fields which create the
open gaps we define the set
\[
 \lb{xj010}
\gX_{p_*}=\rt\{ v\in \ell_{0,p_*}^{per}: \ \ca\hat v_n^0 +\hat
v_n^1\ne 0,\ \hat v_n^0\hat v_n^1=0, \ all\
n\in\N_{p-1},\qq \hat v_p^0\ne0,  p_*\in 2\N\\
 \hat v_n^0\ne0,\ \qqq  \qqq \qq \qqq all\
n\in\N_{p-1},\ \qqq \qqq p_*\ is\ odd \ac\rt\}\rt\}.
\]

\begin{proposition}
\label{Tx} {\it i)  The set $\gX_{p_*}\not=\emptyset$ for any $
p_*\ge2$.

\no ii) If $v\in\gX_{p_*}, p_*\in 2\N$, then
\[
\lb{T3-5}
 z_{k,n}^{b,\pm}(t)=z_{k,n}^{\pm}(0)\pm t\x_n+O(t^2),\qqq
 \x_n=|\hat v_n^1+\hat v_n^0|>0\ \
 as \ \ t\downarrow 0,\qq all \ \
 n\in \N_{k,p}.
\]
iii) If $v\in\gX_{p_*}$ is sufficiently small and $p_*$ is odd, then

\no If  $2|c_k|\ne1$, then each $\p_{k,n}(v)\ne 0$,
$n\in\N_{2p-1}\sm\{p\}$ and $\g_{k,n}^b\ne0 $.

\no If  $2|c_k|=1$, then each $\p_{k,n}(v)=\ca \ne0 \ \ all\ even\
n\in\N_{2p-1}\\ 0 \qqq all\  odd \ n\in\N_{2p-1}\ac$ and
$\g_{k,n}\ne0 $  for any even $n\in\N_{2p-1}$. }
\end{proposition}

\no {\bf Remark.} \er{T3-5} gives the asymptotics of the gap length
$z_{k,n}^{b,+}(t)-z_{k,n}^{b,-}(t)=t2|\hat v_n^j|+O(t^2)$ as $t\to
0$ where $j=0$ or $j=1$. Note that the first term does not depend on
$k\in \Z_N$. If $p_*$ is even, then for large class of potentials
$v\in \gX_p$  all gaps $(z_{k,n}^{b,-}(t),z_{k,n}^{b,+}(t))$ are
open.

We formulate the theorem, motivated by the physical paper of Novikov
\cite{N}.

\begin{theorem}
\lb{T4} Let $v\in\ell_{p_*}^{per}$ and let  $t>0,b\in \R$ be
sufficiently small.

\no i) Let $b=0$. If $N\in 2\N$ and $p$ are coprime numbers, then
$\s_{pp}(H^{0}(t))\ss \cap_{n=1}^{N-1}\s(J_k^{0}(t))$.

\no ii) If  $p>2N$, then the spectrum of
 $H^b(t)$ on the set  $\s(H^{b}(t))\cap ([-\r,-r]\cup [r,\r])$ has multiplicity
 $2$ and satisfies
\begin{multline}
\lb{T4-1}
 \s(H^{b}(t))\cap[r,\r]=\s(J_N^{b}(t))\cap[r,\r]
 =[r,\r]\sm\!\!\bigcup_{2p-1-{p\/N}}^{2p-1}\!\!\g_{N,n}^{b}(t),\ \ r=|2+e^{\frac{i\pi
 }{N}}|, \r=\frac{3+|2+e^{\frac{i\pi}p}|}2,\\
 \s(H^{b}(t))\cap[-\r,-r]=\s(J_N^{b}(t))\cap[-\r,-r]=
 [-\r,-r]\sm\bigcup_{1}^{p\/N}\g_{N,n}^{b}(t).
\end{multline}
 Moreover, if $v\in \gX_{p_*}$, then  each $|\g_{N,n}(t)|>0, n\in \N_{2p-1}$.

\no iii) If $N\not\in 3\N$, then $\s(H^{b}(t))\cap[-r,r]=\es$ for
some $r>0$.

\no iv) If $N\in 3\N$ and $p>2N$, then the spectrum of  $H^b(t)$ on
the set $\s(H^{b}(t))\cap[-r,r]$ has multiplicity $2$ and satisfies
\[
\lb{T4-2} \s(H^{b}(t))\cap[-r,r]=\s(J_{N\/3}^{b}(t))\cap[-r,r]=
[-r,r]\sm\bigcup_{p(1-{1\/N})}^{p(1+{1\/N})}\g_{{N\/3},n}^{b}(t) ,\
\ \ r=|1-e^{\frac{i\pi}N}|,
\]
\[
\lb{T4-3}
|\g_{{N\/3},n}(t)|>0\qq  if \ \ca \qq p_*\in 2\N, \qqq n\in \N_{2p-1}\\
\qq p_* \ is\  odd,\qq  even \ n\in \N_{2p-1} \ac, \qq v\in
\gX_{p_*}.
\]
\end{theorem}

{\bf Remark.} 1) The gaps $\g_{{N},n}^b(t)$ in \er{T4-1} and
$\g_{{N\/3},n}^b(t)$ in \er{T4-2} are also the gaps in the spectrum
of $H^b(t)$. Then we may choose the potentials $v$ such that all
these gaps are open (for wide set of potentials). 2) Due to iii)
$\s(H^b)$ has a gap contained the interval $[-r,r]$

We consider the nanotube in {\bf strong electric fields}. Our
operator has the form  $H^b(t)=H_0^b+t V$, where a coupling constant
$ t\to \iy$. For each $(v_n)_1^{2p}\in \R^{2p}$ there exists a
permutation $\a:\N_{2p}\to \N_{2p}$ such that $h_n=v_{\a(n)}$ and
$h_1\le h_2\le ...\le h_{2p}$. Let $v_n\not=v_j$ for all $n\not=j$,
$n,j\in \N_{2p}$. Defining disjoint intervals
$\mC_n=[th_{n-1}^0,th_{n}^0), h_n^0={h_n+ h_{n+1}\/2}, n\in \N_{2p},
\ \ h_0^0=-\iy, h_{2p+1}^0=\iy$, we obtain the inclusion
$\s(H^b(t))\subset \cup_{n=0}^{2p}\mC_n=\R$. We shall call the set
$\s(H^b(t))\cap \mC_n$ the $n$'th \emph{spectral bands cluster}. Our
goal is to study the asymptotic distribution of eigenvalues in the
$n$'th cluster as $t\to\iy$.

\begin{theorem}
\lb{T44}
 Let $v\in\ell^{per}_{p_*}$, $v_n\not=v_j$ for
all $n\not=j$, $n,j\in \N_{2p}$ and let $c_k=\cos (b+{\pi
k\/N})\not=0$ for some $(k,b)\in\N_N\ts \R$. Let
$v_{\a(n)}<v_{\a(j)}$ for all $n<j$ and some permutation
$\a:\N_{2p}\to \N_{2p}$. If $\wt n=\a^{-1}(n)$ for some
$n\in\N_{2p}$,  then the spectral bands
$\s^{b}_{k,n}(t)=[z_{k,n-1}^{b,+}(t),z_{k,n}^{b,-}(t)]$ satisfy
\[
 \lb{z001l}
 z_{k,\wt n-1}^{b,+}(t)=tv_n-{C_n+O(t^{-1})\/t},\qqq C_n=
 {a_{k,n-1}^2\/v_{n-1}-v_n}+{a_{k,n}^2\/v_{n+1}-v_n},
\]
\[
 \lb{z002l}
 z_{k,\wt n}^{b,-}(t)-z_{k,\wt n-1}^{b,+}(t)={1+O(t^{-1})\/E_n
 t^{2p-1}},\qqq
E_n={1\/2|2c_k|^{p}}\prod_{j\not=n}|v_n-v_j|,
\]
 as $t\to\iy$. Moreover,
\[
\lb{T44-3}
   \s(H^b(t))\cap \mC_{\wt n}(t)=\bigcup_{k=1}^N\s^{b}_{k,\wt n}(t)
 \ss  \lt(v_nt-{\d\/t},v_nt+{\d\/t}\rt),\ \
\d=\max_n{2\/|v_n-v_{n+1}|},
\]
\[
\lb{T44-4}
 \s^{b}_{k,n}(t)\cap\s^{b}_{k',n}(t)=\es,\qqq
if  \qq \ca k\neq k',\ \ \qq b\notin {\pi\/2N}\N\\
 |c_k|\ne |c_{k'}|,\qq b\in {\pi\/2N}\N \ac,
\]
where the spectrum of $H^b(t)$ on $\s^{b}_{k,\wt n}(t)$ has
multiplicity $2$ if $c_k\ne 0$ and $\s^{b}_{k,\wt n}(t)$ is a flat
band if $c_k=0$.
\end{theorem}

{\bf Remark.} 1) Theorems \ref{T3}, 1.4 describe the case $t\to 0$
and Theorem \ref{T44} describe the case $t\to \iy$. These two cases
are quite different, see Fig. 3 and Fig 4.

\begin{figure}[h]
\lb{f0012}
         \centering\includegraphics[clip]{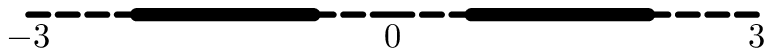}
         \begin{center}
         {\small Fig. 3. Open spectral small gaps for the potential $tV$ as $t\to 0$. }
         \end{center}
\end{figure}

\begin{figure}[h]
\lb{f0011}
         \centering\includegraphics[clip]{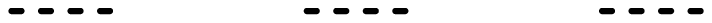}
         \begin{center}
         {\small Fig. 4. Spectral clasters for the potential $tV$ as $t\to \iy$
         for the case $N=4$. }
         \end{center}
\end{figure}

\no 2) The spectral bands cluster $\s(J^b(t))\cap \mC_{\wt n}(t)$ is
a union of $N$ non overlapping bands $\s^{b}_{k,\wt n}(t),k\in
\N_N$, see \er{T44-3}. Recall that if $|c_k|= |c_{k'}|$, then
$J_k^b(t)=J_{k'}^b(t)$.

We present the plan of our paper. In Sect. 2 we prove Theorem
\ref{T1} and \ref{T2}. In the proof Theorem \ref{T1} we use
arguments from \cite{KL}, \cite{KL1}. In the proof Theorem \ref{T2}
we use arguments from \cite{KKu1}. In Sect. 3 we consider the simple
examples for the case $p=1$, in fact, we study unperturbed
Hamiltonians. In Sect. 4 we prove Theorem \ref{T3} -\ref{T44}. In
Sect. 6 we apply some of these methods to analyze the spectral
properties of Shr\"odinger operator on armchair nanotubes.

\section{Proof of Theorems \ref{T1} and \ref{T2}}
\setcounter{equation}{0}

\no {\bf Proof of Theorem \ref{T1}.} i)
 Define an operator $\mJ^b:(\ell^2)^N\to(\ell^2)^N$
 acting on a vector-valued function
 $\p=(\p_n)_{n\in\Z}\in (\ell^2)^N, \p_{2n+1}=(f_{n,0,k})_{k\in \Z_N},\
 \p_{2n}=(f_{n-1,1,k})_{k\in \Z_N}\in\C^N$,  by
\[
 \lb{dJ2}
(\mJ^b \p)_{2n}=((H^b f)_{n,0,k})_{k\in \Z_N},\qqq
  (\mJ^b \p)_{2n+1}=((H^b f)_{n,1,k})_{k\in \Z_N}.
\]
Define a matrix-valued operators $P_n:\C^N\to \C^N$ by
\[
 \lb{dQ2}
 P_{2n+1}h=(V_{n,0,k}h_k)_{k\in \Z_N},\ \
 P_{2n}h=(V_{n,1,k}h_k)_{k\in \Z_N}, \ h=(h_k)_{k\in \Z_N}.
\]
Define the operator $\cS$ in $\C^N$ by $\cS
u=(u_N,u_1,\dots,u_{N-1})^\top$, $u=(u_n)_1^N\in \C^N$. Using
\er{dJ2}, \er{010}, \er{cb},\er{dQ2} and $\cS^*=\cS^{-1},
A=e^{ib}I_N+e^{-ib}\cS^*$ we obtain
$$
 (\mJ^b \p)_{2n+1}=(e^{ib}\cS+e^{-ib})\p_{2n}+\p_{2n+2}+P_{2n+1}\p_{2n+1}
 =A^*\p_{2n}+\p_{2n+2}+P_{2n+1}\p_{2n+1},
$$
$$
 (\mJ^b \p)_{2n}=\p_{2n-1}+(e^{ib}+e^{-ib}\cS^*)\p_{2n+1}+P_{2n}\p_{2n}
 =\p_{2n-1}+A\p_{2n+1}+P_{2n}\p_{2n}.
$$
Finally we rewrite the operator $\mJ^b:(\ell^2)^N\to(\ell^2)^N$ in
the form of the  operator Jacobi by
\[
 \lb{003}
 (\mJ^b \p)_n=A_{n-1}^*\p_{n-1}+A_n\p_{n+1}+P_n\p_n,\qq
 A_{2n}=A=e^{ib}I_N+e^{-ib}\cS^*,\ \ A_{2n+1}=I_N,
\]
$\ \ n\in\Z$, and then
\[
\lb{0031}
 \mJ^b=\left(\begin{array}{ccccccc}
 ... &...&... & ...      &...       &...       &... \\
 ... &A^*  &P_1   & I_N     &0         & 0        &... \\
 ... &0  & I_N& P_2        &A& 0        &... \\
 ... &0  & 0  &A^*&P_3         & I_N     &... \\
 ... &0  & 0  & 0        &I_N      & P_4        &... \\
 ... &0  & 0  & 0        &0         &A^*&... \\
 ... &.  & ...& ...      &...       &...       &... \\
 \end{array}\right).
\]
The matrix-valued function $P_n$ is 2p-periodic. Then the operator
$\mJ^b$ is a 2p-periodic Jacobi operator with $N\ts N$ matrix
-valued coefficients. Note that such operators were considered in
\cite{KKu2}.

The unitary operator $\cS$ has the form $\cS =\sum_1^Ns^k\cP_k$,
where $\cS \wt e_k=s^k\wt e_k$ and $\wt
e_k={1\/N^{1\/2}}(1,s^{-k},s^{-2k},...,s^{-kN+k})^\top$
 is an eigenvector (recall that $s=e^{i{2\pi \/N}}$);
 $\cP_ku=\wt e_k(u,\wt e_k), u=(u_n)_1^N\in \C^N$ is a projector.
Define the operators $\wt \cS\p=(\cS \p_n)_{n\in \Z}$ and
 $\wt \cP_k\p=(\cP_k \p_n)_{n\in \Z}$.
 The operators $\wt \cS$ and $\mJ^b$ commute, then $\mJ^b=\os_1^N
(\mJ^b \wt \cP_k)$. Using \er{003}, \er{0031}  we deduce that $\mJ^b
\wt\cP_k$ is unitarily equivalent to the operator $\mJ_k^b$ given by
\begin{multline}
 \lb{Jk2}
 (\mJ^b_k y)_n=\wt a_{k,n-1}^{*}y_{n-1}+\wt a_{k,n}y_{n+1}+v_ny_n,\qq
 y=(y_n)_{n\in \Z}\in \ell^2,\\
c_k=\cos (b+{\pi k\/N}),\  \ \wt
a_{k,2n}=e^{ib}+e^{-ib}s^{-k}=2e^{-i{\pi k\/N}}c_k,\qq s=e^{i{2\pi
\/N}}, \qq \wt a_{k,2n+1}=1,
\end{multline}
and using Lemma \ref{TAd} we obtain \er{Jk}.

ii) If $c_k=0$, then the Jacobi operator $J_k^b$ has the form
\[
\lb{Jk1}
 J_k^b=\left(\begin{array}{ccccccc}
 ... &...&... & ...      &...       &...       &... \\
 ... &0  &v_1   & 1     &0         & 0        &... \\
 ... &0  & 1& v_2        &0  & 0        &... \\
 ... &0  & 0  &0& v_3         & 1     &... \\
 ... &0  & 0  & 0        &1      & v_4       &... \\
 ... &0  & 0  & 0        &0         &0&... \\
 ... &.  & ...& ...      &...       &...       &... \\
 \end{array}\right)=\os_{n\in \Z}\cJ_n,\qq \cJ_n=\ma v_{2n-1} & 1 \\ 1 & v_{2n} \am.
\]
The eigenvalues of $\cJ_n$    are given by $
z_{n,j}=v_n^++(-1)^j\sqrt{{v_n^-}^2+1}, v_n^\pm={v_{2n-1}\pm
v_{2n}\/2}$ for $(n,j)\in \Z\ts \N_2$, which yields \er{spp}. \BBox

Recall results from \cite{vM} about our $2p$-periodic Jacobi
operator $J(a):\ell^2\to\ell^2$ given by
\[
\lb{j0001}
 (J(a)y)_n=a_{n-1}y_{n-1}+a_ny_{n+1}+v_ny_n,\ \ a_{2n}=a>0,\ \
 a_{2n+1}=1,\ \ n\in\Z, \
 y=(y_n)_{n\in\Z}.
\]
Note that $J_k^b=J(a)$, where $a=2|c_k|, c_k=\cos({\pi k\/n}+b)$.
Introduce fundamental solutions $\vp=(\vp_n(z,a))_{n\in\Z}$ and $
\vt=(\vt_n(z,a))_{n\in \Z} $ for the equation
\[
 \lb{1e}
a_{n-1} y_{n-1}+a_{n}y_{n+1}+v_ny_n=z y_n,\ \ (z,n)\in\C\ts\Z,\qq
a_{2n+1}=1,\ a_{2n}=a,
\]
with initial conditions $\vp_{0}\ev \vt_1\ev 0,\ \vp_1\ev \vt_{0}\ev
1$.  The function $\D={1\/2}(\vp_{2p+1}+\vt_{2p})$ is called the
Lyapunov function for the operator $J(a)$. The functions $\D, \vp_n$
and $\vt_n, n\ge1$ are polynomials of $(z,a,v)\in\C^{2p+2}$. It is
well known that $\s(J(a))=\s_{ac}(J(a))$, where
\begin{multline}
 \lb{j0002}
\s_{ac}(J(a))=\{z\in\R:\ \D(z,a)\in [-1,1]\}=\cup_1^{2p} \s_{n}(a),\
\ \ \ \s_{n}(a)=[z_{n-1}^{+}(a),z_{n}^{-}(a)],
\end{multline}
and $z_{0}^{+}<z_{1}^{-}\le z_{1}^{+}<..\le z_{2p}^{-}$, where
$z_{n}^\pm=z_{n}^\pm(a)$. Note that $\D(z_n^{\pm},a)=(-1)^{p-n}$ for
all $n=0,..,p$. Below we will sometimes write $\s(a,v), J(a,v),..$,
instead of $\s(a), J(a), ..$, when several potentials are being
dealt with. Recall that  the $2p\ts 2p$ matrix $K(\t,a)$ is given by
\[
\lb{dL}
 K(\t,a)=K^0(\t,a)+B,\qq
 K^0(\t,a)=\left(\begin{array}{ccccc} 0 & 1 & 0 & ... & {a\/\t} \\
                                   1 & 0 & a & ... & 0 \\
                                  0 &  a & 0 & ... & 0 \\
                                  ... & ... & ... & ... & ... \\
                                  \t a & 0 & ... & 1 & 0
                                  \end{array}\right),\
 B=\diag (v_n)_1^{2p},
\]
where $\t\in \S^1=\{\t\in\C:|\t|=1\}$. Fix $a, \f\in [0,2\pi]$, then
eigenvalues of $K(e^{i\f},a)$ are all zeros of the polynomial
$\D(z,a)-\cos \f$. Then the fundamental solutions $\vp_{k,n},
\vt_{k,n}$, the Lyapunov function and the spectral bands
$\s_{k,n}^b$ for the operator $J_k^b$ satisfy (see also \er{sJk})
\[\lb{pJ00}
\vp_{k,n}=\vp_{n}(z,a),\qq \vt_{k,n}=\vt_{n}(z,a),\qq \D_k=\D(z,a)
\qq z_{k,n}^{b,\pm}=z_n^{\pm}(a),
\]
\[
 \lb{pJ}
\s(J_k^b)=\s_{ac}(J_k^b)=\{z\in\R:\ \D_k(z)\in [-1,1]\}=\cup_1^{2p}
\s_{k,n}^b, \qq \s_{k,n}^b=[z_{k,n-1}^{b,+},z_{k,n}^{b,-}],
\]

\no {\bf Proof of Theorem \ref{T2}.}
 Let $a=2|c_k|\to 0$. We consider the matrix
$ K(\t,a)$ as $a\to 0,\t\in \S^1=\{\t\in\C:|\t|=1\}$. If $a=0$, then
we get $K(\t,0)=\os_1^p\cJ_n$, where $\cJ_n$ is given by \er{Jk1}.
Let $\l_1\le \l_2\le ....\l_{2p}$ be the eigenvalues of $K(\t,0)$.
The endpoints $z_{n-1}^{+}(a),z_{n}^{-}(a)$ of the spectral bands
$\s_{n}(a)=[z_{n-1}^{+}(a),z_{n}^{-}(a)]$ of the operator $J(a)$ are
the eigenvalues  of  $K(\pm1,a,v)$. By the perturbation theory
\cite{RS}, they are analytic function from $a$ and if $a\to 0$, then
the spectral bands converge to the set $\{\l_1,\l_2,....\l_{2p}\}$.
The number of spectral bands converging to $\l_n$ coincides with the
multiplicity of $\l_n$ as $a\to 0$. In particular, if some $\l_n,
n\in \N_{2p}$ is simple, then $\s_n(a)\to \{\l_n\}$.

Recall that the monodromy matrix $M_{2p}$ for the operator $J(a)$ is
given by
\begin{multline}
 \lb{m1}
 M_{2p}(z)=\ma \!\!\!\vt_{2p} & \vp_{2p} \!\!\!\\
           \!\!\!\vt_{2p+1} & \vp_{2p+1}\!\!\!
       \am=T_{p}.. T_2T_1,\ \ \\
 T_n=
{1\/a} \ma \!\!\!0 & a \!\!\! \\ \!\!\!-1 & z-v_{2n+1} \!\!\!\am
 \ma\!\!\!0 & 1\!\!\! \\ \!\!\!-a & z-v_{2n}\!\!\! \am=
 \ma -a&  z-v_{2n-1}\\ v_{2n}-z& \f_n/a\am,
\end{multline}
where $ \f_n=(z-v_{2n})(z-v_{2n-1})-1$. Let
$$
X_n=ET_n E_1=\ma \f_n&  v_{2n}-z\\ z-v_{2n-1}& -1\am,\ E=\ma
0&a\\1&0\am,\ E_1=\ma 0&{1\/a}\\1&0\am,
$$
$$
A=(EE_1)^{-1}=\ma a&0\\0&{1\/a}\am={1\/a}A_1,\qq A_1=\ma
a^2&0\\0&1\am.
$$
Then $ M_{2p}=E^{-1} X_{p}A X_{p-1}A.. AX_1E_1^{-1}, $ which yields
the Lyapunov function $\D$ given by
$$
2\D=\Tr M_{2p}=\Tr X_{p}A X_{p-1}A.. AX_1A= {1\/a^p}\Tr X_{p}A_1
X_{p-1}A_1.. AX_1A_1={1\/a^p}\sum_{n=0}^{p}a^{2n}\F_n(z),
$$
and
\[
 \lb{b001}
 \D(z,a)={\F_0(z)+a^2\F(z,a^2)\/2a^{p}},\qq
\F_0=\prod_{n=1}^{2p}(z-\l_n), \qq
\F(z,t)=\sum_{n=1}^{p}t^{n-1}\F_n(z),
\]
for some polynomials $\F_n$.  By the perturbation theory (see
\cite{RS}), the endpoints $z_+, z_-$ of the spectral band
$\s_{s}(a)=[z_{s-1}^{+}(a),z_{s}^{-}(a)]=[z_+,z_-]$ are analytic
functions in some disk $\{a\in \C:|a|<\ve \},\ve >0$ and satisfy the
equation $\D(z_\pm,a)=\mp(-1)^s$, which has the form
\[
 \lb{b002}
 \F_0(z_{\pm})+a^2\F(z_{\pm},a^2)=\mp(-1)^s 2a^p.
\]
Moreover, they satisfy $z_{\pm}(a)=\l_s+O(a^2)$ as $a\to 0$ at
$p\ge2$ (see the case $p=1$ in Sect. 3).

Let $\l_s$ be a simple eigenvalue for some  $s\in\N_{2p}$. The
differentiation of \er{b002} yields
\[
\lb{di1}
 z_{\pm}'(a)\O+\pa_a(a^2\F(z_{\pm},a^2))=\mp p(-1)^s 2a^{p-1},\qq
 \O(z,a)=\pa_z (\F_0(z)+a^2\F(z,a^2)).
\]
The differentiation of \er{b002} $r\in [1,p]$ times  yields
\[
\lb{dir}
z_{\pm}^{(r)}\O(z_\pm)+G_r(z_{\pm}^{(r-1)},...,z_{\pm},a)=\mp{p!\/(p-r)!}(-1)^s
2a^{p-r},
\]
for some polynomial $G_r$. Then at $a=0$ this gives
\[
\lb{dir0}
z_{\pm}^{(r)}(0)(-1)^s\L_s+G_r(z_{\pm}^{(r-1)}(0),...,z_{\pm}(0),0)=\mp{p!\/(p-r)!}(-1)^s
2a^{p-r}|_{a=0}.
\]
Thus we obtain $ z_{\pm}^{(2r+1)}(0)=0$ for all $2r+1<p$, since the
polynomial $\F=\F(z,a^2)$. Moreover, using $z_{-}(0)=z_{+}(0)$  we
obtain $z_{-}^{(r)}(0)=z_{+}^{(r)}(0)$ for all $r<p$.

Consider the case $r=p$. Identity \er{dir} implies
\[
\lb{dir0p}
z_{\pm}^{(p)}(0)(-1)^s\L_s+G_p(z_+^{(p-1)}(0),...,z_+(0),0)=\mp p!2
(-1)^s,
\]
which yields $z_{\pm}^{(p)}(0)={p!\/\L_s}(C_p\mp 2)$ for some
constant $C_p\in \R$. Using this and
$\s_{s}(a)=[z_{s-1}^{+}(a),z_{s}^{-}(a)]=[z_+,z_-]$ and \er{j0002},
\er{pJ00} we obtain \er{T2-2}, \er{T2-3}. \BBox

\begin{lemma}
\lb{TAd} Let a Jacobi operator $J:\ell^2\to \ell^2$ is given by
\[
\lb{TAd-1}
 (J y)_n=a_{n-1}^{*}y_{n-1}+a_{n}y_{n+1}+v_ny_n,\qq y=(y_n)_{n\in
\Z}\in\ell^2, \ \ a_{n+p}=a_n\in\C,\ \ v_n\in \R,\ \
\]
$ n\in\Z,$ for some $p\ge 1$. Then
\[
\lb{TAd-3} \P^*J\P=J^+,\qq
(J^+y)_n=|a_{n-1}|y_{n-1}+|a_{n}|y_{n+1}+v_ny_n,
\]
where the unitary diagonal operator $\P$ is given by
\[
\lb{TAd-2} \P y=(u_ny_n)_{n\in\Z},\ \ \ u_n=\prod_1^{n}\ol \ve_{j},\
n\ge 0,\ \ \ u_n=\prod_1^{n}\ol \ve_{j},\ n<0,\ \
\ve_n=\ca {a_n\/|a_n|} & if \ a_n\ne 0\\
                        1 & if \ a_n= 0 \ac .
\]
\end{lemma}
{\bf Proof.} Direct calculations give \er{TAd-3}. \BBox

\section{
Example for the case $p=1$} \setcounter{equation}{0}

In this section we consider the Jacobi operator $J_k^b, k\in \Z_N$
given by
\[
\lb{31}
 J_k^b=\left(\begin{array}{ccccccc}
 ... &...&... & ...      &...       &...       &... \\
 ... &a  & v  & 1     &0         & 0        &... \\
 ... &0  & 1& -v        &a& 0        &... \\
 ... &0  & 0  &a& v         & 1     &... \\
 ... &0  & 0  & 0        &1      & -v       &... \\
 ... &0  & 0  & 0        &0         &a&... \\
 ... &.  & ...& ...      &...       &...       &... \\
 \end{array}\right), \  a=2|c_k|,\qq v=v_{2n+1}=-v_{2n}\in \R,\ \  n\in\Z,
\]
i.e., the case $p=1$. The monodromy matrix $M_2$ satisfies (see
\er{m1})
\[
\lb{M2}
M_2(z)=  \ma \!\!\!\vt_{2} & \vp_{2} \!\!\!\\
           \!\!\!\vt_{3} & \vp_{3}\!\!\!
       \am =\ma 0 & 1 \\ -{1\/a} & {z+v\/a} \am \ma 0 & 1 \\ -a &
       z-v
\am =\ma -a & z-v \\ -z-v & {z^2-v^2-1\/a} \am .
\]
Let $\D^0={\Tr M_2\/2}={z^2-v^2-5\/4}$ be the Lyapunov function for
the case $a=1$. This yields
\[
\lb{D2} \D_k={\Tr M_2\/2}={z^2-v^2-4c_k^2-1\/4|c_k|}=
{\D^0+s_k^2\/|c_k|},\qq c_k=\cos (b+{\pi k\/N}).
\]
The periodic eigenvalues $z_{k,0}^{b,\pm}$ satisfy the equation
$\D_k(z)=1$ and anti-periodic eigenvalues $z_{k,1}^{b,\pm}$ satisfy
the equation $\D_k(z)=-1$ and they are given by
\[
\lb{eib}
 z_{k,0}^{b,\mp}=\pm \sqrt{v^2+(2|c_k|+1)^2},\qqq
z_{k,1}^{b,\pm}=\pm \sqrt{v^2+(2|c_k|-1)^2}.
\]
The spectrum of $J_k^b$ has the form
\[
\lb{eib2} \s(J_k^b)=[z_{k,0}^{b,+},z_{k,1}^{b,-}]\cup
[z_{k,1}^{b,+},z_{k,0}^{b,-}]=[z_{k,0}^{b,+},z_{k,1}^{b,-}]\sm
\g_{k,1},\qq
\]
where $\g_{k,1}^b=(z_{k,1}^{b,-},z_{k,1}^{b,+})$ is a gap. Note that
\[
\  \g_{k,1}^b=(z_{k,1}^{b,-},z_{k,1}^{b,+})\ne \es,\qq if \qq
|c_k|\ne {1\/2}.
\]
Let $c_k\to 0$. Then \er{T2-2},\er{T2-3} yield
\begin{multline}
|s_1^b|=z_{k,1}^{b,-}-z_{k,0}^{b,+}=-{4|c_k|\/w}+O(c_k^2),\qq \
w=\sqrt{1+v^2},\\ z_{k,1}^{b,-}=-w+{2|c_k|\/w}+O(c_k^2), \qq
z_{k,0}^{b,+}=-w-{2|c_k|\/w}+O(c_k^2).
\end{multline}

\no {\bf 1. The operator $H^0$, no magnetic field, $b=0$}. In this
case using \er{eib}, \er{eib2}, we obtain
\[
z_{k,0}^{0,+}<z_{0,0}^{0,+}, \qq z_{0,0}^{0,\pm}=\pm
\sqrt{v^2+9},\qq \g_{k,1}=(z_{k,1}^{0,-},z_{k,1}^{0,+})
\ca =\es    & if \  k\in\{{N\/3}, {2N\/3}\}\\
  \ne \es    & if \ k \notin \{{N\/3}, {2N\/3}\}     \ac ,
\]
and then
\[
\s(H^0)=\s_{ac}(H^0)\cup \s_{pp}(H^0), \qq
\s_{pp}(H^0)=\ca\es    & if \  {N\/2}\notin \N\\
    \{\pm \sqrt{1+v^2}\}  & if \ {N\/2}\in \N   \ac,
\]
\[
\s_{ac}(H^0)=[z_{0,0}^{0,+},z_{0,0}^{0,-}]\sm \g(H^0),\qq \g(H^0)=
\ca \es    & if \  {N\/3}\in \N\\
  (z_{m,0}^{0,-},z_{m,0}^{0,+})\ne \es   & if  {N\/3}\notin \N    \ac ,
\]
for some $m\in \Z_N$, where roughly speaking $m  \sim {N\/3}$.

\no {\bf 2. Magnetic field, $b\ne 0$}. Using i) of Theorem \ref{T1}
we obtain $\s(H^{b+{\pi\/N}})=\s(H^b ), \qq b\in \R$. Then we need
to consider only the case $b\in (0,{\pi\/N})$. Using \ref{T1} we
obtain
\[
\s(H^b)=\s_{ac}(H^b)\cup \s_{pp}(H^b), \qq
\s_{pp}(H^b)=\ca\es    & if \  c_k\ne 0, all\ k\in \Z_N\\
    \{\pm \sqrt{1+v^2}\}  & if \ c_k\ne 0, some \ k\in \Z_N
    \ac,
\]
\[
\s_{ac}(H^b)=[z_{0}^{b,+},z_{0}^{b,-}]\sm \g(H^b),\ \g(H^b)=
(z_{1}^{b,-},z_{1}^{b,+}),
\]
where $\g(H^b)$ is the gap in the spectrum of $H^b$ and
\[
z_{0}^{b,+}=\ca z_{0,0}^{b,+}    & if  \qq b\le{\pi\/2N}\\
    z_{N-1,0}^{b,+}  & if  \qq b>{\pi\/2N}  \ac,
\]
and
\[
\g(H^b)=(z_1^{b,-},z_1^{b,+}),\qq  z_1^{b,\pm}=\pm
\sqrt{v^2+(2|c_k|-1)^2}, \  \ for \ some \ k\in \Z_N,
\]
where roughly speaking $2|c_k|  \sim 1$.

\section{Proof of Theorems \ref{T3}-\ref{T44}.}
\setcounter{equation}{0}

{\bf Proof of Theorem \ref{T3}.} In order to determine the
asymptotics \er{T3-1} we need the following fact from the
perturbation theory \cite{RS}: Let $A(t)=A_0+tA_1, t\in\R$, where
$A_0=A_0^*$, $A_1=A_1^*$ are operators in $\C^{2p}$. Let $\m$ be an
eigenvalue of $A_0$ of multiplicity $2$ and let $h^\pm$ be the
corresponding orthonormalized eigenvectors. Then there are $2$
functions $\m_\pm(t)$ analytic in a neighborhood of $0$, which are
all the eigenvalues. Moreover, $\m_\pm(t)=\m+\m_\pm'(0)t+O(t^2)$ as
$t\to 0$, where $\m_\pm'(0)$ are the eigenvalues of $P^*A_1P$ and
$P=(h^-,h^+)$ is the $2p\ts 2$ matrix.

We determine the asymptotics \er{T3-1} of $z^{b,\pm}_{k,n}(t)$ for
$k\in \N_p, n\not=0,p,2p$, the proof of other cases is similar. We
apply the perturbation theory to the operator $K(\pm
1,a,tv)=K^0(\t,a)+tB$ as $t\to 0$, where $K$ is given by \er{dL} and
$a=2|c_k|$. Recall that  $z^{b,\pm}_{k,n}(t)$  are eigenvalues of
$K(\pm 1,a,tv)$, (see \er{j0001}-\er{pJ00}). The operators $K^0(\pm
1,a)$ has eigenvalues
$z_{k,n}^{b,+}(0)=z_{k,n}^{b,-}(0)=\l_n^{\pm}(a)$ (with multiplicity
2) and the corresponding eigenvectors
\[ \lb{j1111}
 Z_{k,n}^\pm=Z_n^\pm(a),\ \ \qqq \ n\in\Z_{2p-1},
\]
see Corollary \ref{b012} and \er{pJ00}, Then by this fact, the
derivatives $(z_{k,n}^\pm)'(0)$ are eigenvalues of the $2\ts
2$-matrix $P_{k,n}^*BP_{k,n}$, where $P_{k,n}=(Z_{k,n}^+,Z_{k,n}^-)$
is the $p\ts 2$-matrix. Define the vectors
\[
 \lb{c001}
 F_n=(2p)^{-1}(f_j)_{1}^{2p},\ \
 \ f_{2j+1}=\t_{n}^{2j}e^{2i\arg(2|c_k|+\t_{n})},\ \ f_{2j}=\t_n^{2j},\
 \ \ \t_n=e^{\frac{i\pi n}p},\ \  j\in \N_p.
\]
Let $\wt v_n=\langle v,F_n\rangle$, $n\in\N_p$ and $\wt v_{p+n}=\wt
v_{p-n}$, $n\in\N_{p-1}$. Using \er{j1111}, Corollary \ref{b012},
\er{j1112}
we obtain
$$
 P_{k,n}^*BP_{k,n}=\ma \Tr B &\langle b,F_n\rangle\\
 \langle F_n,b\rangle& \Tr B \am=\ma 0 & \wt v_n \\ \ol{\wt v_n} & 0\am,\qq
 where \qq B=\diag(v_j)_{1}^{2p}.
$$
The eigenvalues of the last matrix have the form  $\pm|\wt v_n|$,
which yields $(z_{k,n}^{\pm})'(0)=\pm|\wt v_n|$. Recall that the
orthogonal basis in $\C^p$ is given by
$e_n={1\/2p}(\t_n^{2j})_{j=1}^p$, \ $n\in \N_p$, where $
\t_n=e^{i\frac{\pi n}p}$ and the vectors $v^0=(v_{2n})_{n=1}^p$ and
$v^1=(v_{2n-1})_{n=1}^p$, $\hat v^j_n=\langle v^j,e_n\rangle$, $n\in
\N_p$, $j=0,1$. Then \er{c001} gives $\wt v_n=\hat
v_n^0+e^{2i\arg(a+\t_n)}\hat v_n^1$ and we obtain \er{T3-1}.


Let $\cS(u_1,..,u_{p})=(u_p,u_1,..,u_{p-1})$ be a shift operator. If
$p_*$ is odd, then $p=p_*$ and $v^1=S^{\frac{p+1}2}v^0$ and using
\er{j010}, we obtain
$$
 \hat
 v_n^1=\langle v^1,e_n\rangle=\langle S^{\frac{p+1}2}v^0,e_n\rangle=
 \langle v^0,S^{-\frac{p+1}2}e_n\rangle=
 \langle v^0,\t_n^{p+1}e_n\rangle= \t_n^{-p-1}\hat v_n^0=\t_n^{p-1}\hat
 v_n^0,
$$
since $\t_n^{p}=\t_n^{-p}$. Then we get
$$
 \hat v_n^0+e^{2i\arg(2|c_k|+\t_n)}\hat v_n^1=
 \hat v_n^0(1+\t_n^{p-1}e^{2i\arg(2|c_k|+\t_n)}).
$$
Simple calculations gives: if $2|c_k|\not=1$ and $n\in\N_{k,p}$,
then $
 1+\t_n^{p-1}e^{2i\arg(2|c_k|+\t_n)}\ne 0$, and if $2|c_k|=1$, then $
 1+\t_n^{p-1}e^{2i\arg(2|c_k|+\t_n)}=\ca\ne0\ n\ is\ even\\0,\ n\ is\
 odd\ac.
$ \BBox

{\no \bf Proof of Proposition \ref{Tx}.} i) Consider the case $p_*$
is even. Denote $\ol{z}=(\ol z_n)_1^p$ for $z=(z_n)_1^p\in \C^p$.
Using \er{j010}, we obtain $e_{p-n}=\ol{e_n}$, $n\in\N_{p-1}$ and
$e_p=(2p)^{-1}(1,..,1)^{\top}\in\R^p$. If
$v^1=\sum_{n=1}^{p-1}\a_ne_n+\a_pe_p$,\ \ $\ol{\a_n}=\a_{p-n}\ne0$,
$n\in\N_{p-1}$, $0\ne\a_p\in\R$, then $v^1\in\R^p$ and $\hat
v^1_n=\a_n\ne0$, $n\in\N_p$, since $\{e_n\}_1^p$ is orthogonal basis
in $\C^p$. Consider $v^0=-\hat v_p^1e_p$, then $v^0\in\R^p$, since
$e_p\in\R^p$ and $\hat v^1_p=\a_p\in\R$. Also $\hat
v_p^0=-\a_p\ne0$. Then the vector
$v=(v_1^0,v_1^1,..,v_p^0,v_p^1)\in\gX_{p_*}$, since
$\sum_{n=1}^p(v_n^1+v_n^0)=\hat v_p^0+\hat v_p^1=0$. Then
$\gX_{p_*}\ne\es$. The proof of the case of odd $p_*$ is similar.
The statements ii) and iii) follows from Theorem \ref{T3},
\er{T3-5a}. \BBox

 {\no \bf Proof of Theorem \ref{T4}.}
i) Using  \er{tp000}, we obtain that
$(1-\d,1+\d)\cup(-1-\d,-1+\d)\ss\s(J_k^{0}(0))$ for any
$k\in\N_{N-1}\sm\{\frac N2\}$ and for some $\d>0$. If $k=\frac N2$
then we obtain $\s(J_k^{0}(t))=\s_{pp}(H^{b}(t)$. Moreover, we have
that $\s_{pp}(H^{b}(t))\in((1-\d,1+\d)\cup(-1-\d,-1+\d))$ for small
$t$, then in order to prove i) we have to show that there are no
gaps  in small neighborhood of $\{\pm1\}$, i.e.  we need to show
that $z_{k,n}^{b,\pm}\not\in\{\pm1\}$, i.e.
$$
 |2|c_k|+\t_n|\not=1,\ \ n\in\N_p
$$
or
$$
 \lt|2\cos\frac{k\pi}N+\cos\frac{n\pi}p+i\sin\frac{n\pi}p\rt|=1+
 4\cos\frac{k\pi}N\lt(\cos\frac{n\pi}p+\cos\frac{k\pi}N\rt)\ne1
$$
or
\[
 \lb{f010}
 \cos\frac{n\pi}p+\cos\frac{k\pi}N\ne0,
\]
since $\cos\frac{k\pi}N\ne0$ for $k\ne\frac{N}2$. The identity
\er{f010} holds true, since $p$ and $N$ are coprime.

ii) Consider the case $\s(H^{b}(t))\cap[-\r,-r]$
the proof of other cases is similar. Theorem \ref{T3} gives
\[
 \lb{tp000}
 \s(J_k^0(0))=[-2|c_k|-1,-|2|c_k|-1|]\cup[|2|c_k|-1|,2|c_k|+1],\ \
 k\in\Z_N,
\]
which yields $ \s(J_N^0(0))=[-3,-1]\cup[1,3]$ and
$$
 [-\r-\d,-r+\d]\ss J_{N}^0(0),\ \ \ \
 [-\r-\d,-r+\d]\cap J_n^0(0)=\es,\ \
 k\in\N_{N-1}
$$
for some small $\d>0$ (see \er{Jk} and before \er{T4-1}). Then the
spectrum in $\s(J^{b}(t))\cap[-\r,-r]$ has multiplicity $2$ for all
sufficiently small $t$ and $b$. Also, using \er{T3-3}, we obtain
$z_{N,n}^{0}(0)\in[-\r,-r]$, $1\le n\le\frac pN$ and
$z_{N,n}^{0}(0)\not\in[-\r,-r]$ for $n>\frac pN$, which
yield\er{T4-1}. The inequality $|\g_{N,n}(t)|>0$ follows from
Proposition \ref{Tx}.

iii) follows from \er{tp000}, since $\s(J_k^0(0))\cap[-r,r]=\es$ for
any $k$ and sufficiently small $r>0$.

 The proof of iv) is similar to the proof of ii).
 \BBox

{\no \bf Proof of Theorem \ref{T44}.} Recall that $ (J_k^b(t)
y)_n=a_{n-1}y_{n-1}+a_{n}y_{n+1}+tv_ny_n,\ y=(y_n)_{n\in \Z}\in
\ell^2,   \
 n\in\Z,
$ where $a_{2n}=2|c_k|,\ a_{2n+1}=1$. Using \er{dL} we obtain
$$
 K_{k}(\t,a,tv)=t(B+\ve K^0(\t,a)),\qq B=\diag(v_j)_{1}^{2p}\ as \
                                   \ \ve={1\/t}\to 0,\qq a=2|c_k|.
$$
Then the perturbation theory \cite{RS} for $B+\ve K^0(\t,a)$ gives
$$
 {\l_n(t)}=t(v_n+\ve u_{n,n}+\a_n\ve^2+O(\ve^3)),\qqq
 \a_n=-\sum_{j\not=n}{u_{n,j}u_{j,n}\/v_j-v_n},\qqq
 u_{j,n}=(e_j^0,K^0(\t)e_n^0),
$$
where $Be_j^0=v_je_j^0$ and the vector
$e_j^0=(\d_{j,n})_{n=1}^{2p}\in \C^{2p}$. The definition of
$u_{j,n}$ yields
$$
 u_{n-1,n}=u_{n,n-1}=a_{n-1},\ \ u_{n+1,n}=u_{n,n+1}=a_{n},\ \ and
 \ \
 u_{n,j}=0\ \ if\ \ |j-n|\not=1.
$$
These imply \er{z001l}, since $z_{k,n}^{b,\pm}(t)$ are eigenvalues
of $K_{k}(\pm 1,a,tv)$.

 We show \er{z002l} for the case $v_1<..<v_{2p}$, the proof
of other cases is similar. Using the identity $
2\D_k(z,t)\ev2\D(z,a,tv)=w^{-1}\det(zI_{2p}-K(i,a,tv))$
 (see reasoning between \er{j0001} and \er{pJ}),
 where $w=\prod_{1}^{2p}a_{n}=|2c_k|^{p}\ve^{2p}$, we obtain
\[
 \lb{b100}
2\D_k(z,t)=w^{-1}\det(z\ve I_{2p}-B+\ve K^0(i,a))
 =\frac{F_0(\l)+\ve F(\l,\ve)}{|2c_k|^{p} \ \ve^{2p}},\qq
 \l=\frac zt=z\ve,
\]
where $F_0(\l)=\det(\l I_{2p}-B)=\prod_{j=1}^{2p}(\l-v_j)$ and $F$
is some polynomial of two variables  $\l,\ve$.

   Let $\l_+(\ve)=z_{k,n-1}^{b,+}(t)/t,
   \l_-(\ve)=z_{k,n}^{b,-}(t)/t$ for some $n\in \N_{2p}$.
   These $\l_\pm(\ve)$ are the solutions of the equation $F(\l_\pm,\ve)=\pm 1$,
   where $F(\l,\ve)=\D_k(z,t)$ and \er{z001l}
yields $\l_{\pm}(\ve)=v_n+O(\ve^2)$ as $\ve\to0$. By the
perturbation theory \cite{RS}, the functions $\l_\pm(\ve)$ are
analytic in some disk $\{|\ve|<r\},r>0$. Now we repeat the arguments
from the proof of Theorem \ref{T2} after \er{b002}. Differentiating
\er{b100} $2p$ times we obtain
$$
 (\l_+)^{(j)}(0)=(\l_-)^{(j)}(0),\ \ j<2p,\ \
 (\l_+)^{(2p)}(0)-(\l_-)^{(2p)}(0)={(2p)!\/E_n},\qq
$$
i.e.,
$$
 |\l_+(\ve)-\l_-(\ve)|={\ve^{2p}\/E_n}+O(\ve^{2p+1})\qq as \ \ \
 \ve\to0,\qq
  E_n={1\/2|2c_k|^{p}}\prod_{j\not=n}(v_n-v_j),
$$
which yields \er{z002l}, since $\l_+(\ve)=z_{k,n-1}^{b,+}(t)/t,
   \l_-(\ve)=z_{k,n}^{b,-}(t)/t$.
Using \er{z001l}, we obtain \er{T44-3}.

If $|c_{k}|\ne |c_{k'}|$,  then \er{z002l} implies
$$
 \s^{b}_{k,n}(t)\cap\s^{b}_{k',n}(t)=
 [z_{k,n-1}^{b,+}(t),z_{k,n}^{b,-}(t)]\cap[z_{k',n-1}^{b,+}(t),z_{k',n}^{b,-}(t)]=\es
$$
for sufficiently large $t$. This yields  \er{T44-4} for the second
case. If $k\ne k'$ for $k,k'\in\N_{N}$ and $b\notin {\pi \/2N}\N$,
then $|c_{k}|\ne |c_{k'}|$ and we obtain \er{T44-4} for the first
case.

Using \er{T44-4}, we obtain $\s(J_k^b(t))\cap\s(J_{k'}^b(t))=\es$,
$k\ne k'$. Then $\s(J_k^b(t))$ has multiplicity $2$ and
$\s_{k,n}^b(t)$ has multiplicity $2$. \BBox

\section{Armchair nanotube.}
\setcounter{equation}{0}
\begin{figure}[h]
         \centering\includegraphics[clip]{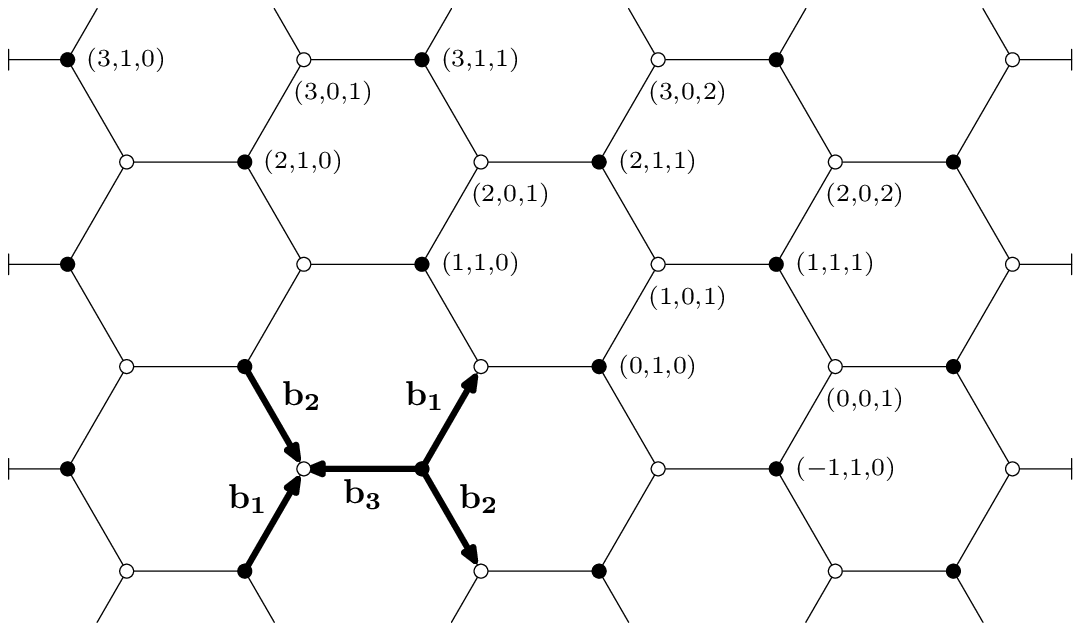}
         \begin{center}
         {\small Fig 3. A piece of armchair nanotube. }
         \end{center}
         \lb{f001arm}
\end{figure}
\begin{figure}[h]
         \centering\includegraphics[clip]{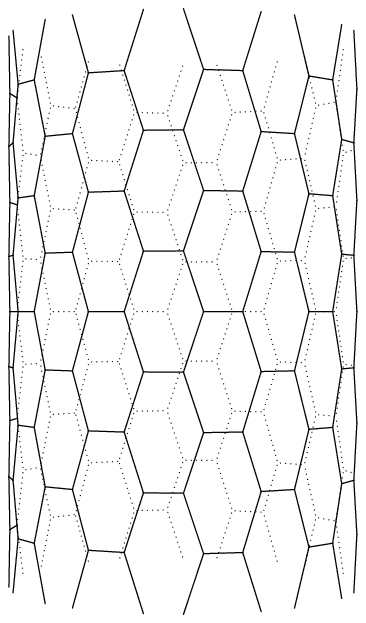}
         \begin{center}
         {\small Fig 4. 3D model of armchair nanotube. }
         \end{center}
         \lb{f001arm}
\end{figure}
We consider the Schr\"odinger operator $H^b(v)$ with a real periodic
potential $v$ on the armchair nanotube $\G\ss\R^3$ in a uniform
magnetic field $\mB=B(0,0,1)\in \R^3$, $B\in\R$. Our model armchair
nanotube $\G$ is a graph (see Fig. \ref{f001arm}) embedded in $\R^3$
oriented in the $z$-direction ${\bf e}_0$. $\G$ is a set of vertices
(atoms) ${\bf r}_{\o}$ connecting by bonds (edges) and
\[
\G=\cup_{\o\in \cZ} {\bf r}_{\o},\ \  \o=(n,j,k)\in \cZ=\Z\ts
\{0,1\}\ts \Z_N, \qq \Z_N=\Z/(N\Z),
\]
where $N$ is a number of vertices in any ring of nanotube. The
detail information about $3D$ coordinates of $r_{\o}$ and about
constants $b_j$ see in Appendix.

Introduce the Hilbert space $\ell^2(\G)$ of functions
$f=(f_\o)_{\o\in \cZ}$ on $\G$ equipped with the norm
$\|f\|_{\ell^2(\G)}^2=\sum_{\o\in \cZ} |f_\o|^2 $. The tight-binding
Hamiltonian $ H^{b}$ (where $ b=(b_1,b_2,b_3)$) on the nanotube $\G$
has the form $ H^{b}= H_0^{b}+V$ on $\ell^2(\G)$, where $H_0^b$ is
the Hamiltonian of the nanotube in the magnetic field and is given
by
\begin{multline}
 \lb{010arm}
 ( H_0^{b} f)_{n,0,k}=e^{ib_2}f_{n+1,1,k}+e^{ib_1}f_{n-1,1,k-1}+
 e^{ib_3}f_{n,1,k},\\
 ( H_0^{b} f)_{n,1,k}=e^{-ib_1}f_{n+1,0,k+1}+e^{-ib_2}f_{n-1,0,k}+
 e^{-ib_3}f_{n,0,k},\qq f=(f_\o)_{\o\in \cZ},\\
 \o=(n,j,k)\in \Z\ts \{0,1\}\ts \Z_{N}
\end{multline}
and the operator $V$ corresponding to the  external electric
potential
 is given by
\[
\lb{cbarm} (Vf)_\o=V_\o f_\o,\qq where \qq V_{n,0,k}=v_{2n}, \qq
V_{n,1,k}=v_{2n+1},\qq k\in\Z_N, \qq v=(v_n)_{n\in\Z}\in\ell^\iy.
\]

{\no\bf 1. The operator $H^b$ is an orthogonal sum of Jacobi
operators.}
\begin{theorem}
\label{T1arm} Let $v=(v_n)_{n\in \Z}\in \ell^\iy$. Then the operator
$ H^{b}$ is unitarily equivalent to the operator $\os_1^N J_k^{b}$,
where $ J_k^{b}$ is a Jacobi operator, acting on
$\ell^2(\Z)\os\ell^2(\Z)$ and given by
\begin{multline}
 \lb{Jkarm}
 ( J_k^{b} y)_n=ay_{n-1}+a^*y_{n+1}+d_ny_n,\qq
 y=(y_n)_{n\in \Z}\in \ell^2\os\ell^2,\\
  a\ev a_{k}=\ma 0 & e^{ib_1}s^k\\e^{-ib_2} &0 \am,\ \ s=e^{\frac{2\pi
  i}N},  \qq d_n\ev \ma v_{2n} & e^{ib_3}\\e^{-ib_3}&v_{2n+1} \am,\qq
 n\in\Z.
\end{multline}
Each $ J_k^{b}$ has absolutely continuous spectrum.
\end{theorem}

\no {\bf Proof of Theorem \ref{T1arm}.} We give compressed Proof
because this one is similar to the Proof of Theorem \ref{T1}.
 Define the operator $\mJ^{b}:(\ell^2)^{2N}\to(\ell^2)^{2N}$
 acting on a vector-valued function
 $\p=(\p_n)_{n\in\Z}\in (\ell^2)^{2N}$,
 $\p_{n}=(f_{n,0,k},f_{n,1,k})_{k\in \Z_N}^{\top}\in\C^{2N}$,  by
\[
 \lb{dJ2arm}
(\mJ^{b} \p)_{n}=(( H^{b} f)_{n,0,k},( H^{b} f)_{n,1,k})_{k\in
\Z_N}^{\top}.
\]
Define the operator $\cS$ in $\C^N$ by $\cS
u=(u_N,u_1,\dots,u_{N-1})^\top$, $u=(u_n)_1^N\in \C^N$. Using
\er{dJ2arm}, \er{010arm}, \er{cbarm} and $\cS^*=\cS^{-1}$ we obtain
\[\lb{003arm}
 (\mJ^{b} \p)_{n}=A\p_{n-1}+
 A^*\p_{n+1}+C_n\p_{n},\ \ where
\]
\[\lb{0031arm}
 A=\ma 0 & e^{ib_1}\cS \\ e^{-ib_2}I_N &0 \am,\qq
 C_n=\ma v_{2n}I_N & e^{ib_3}I_N \\ e^{-ib_3}I_N & v_{2n+1}I_N \am.
\]
The unitary operator $\cS$ has the form $\cS =\sum_1^Ns^k\cP_k$,
where
$$
 \cS \wt e_k=s^k\wt e_k,\ \ \ \wt
 e_k={1\/N^{1\/2}}(1,s^{-k},s^{-2k},...,s^{-kN+k})^\top
$$
 is an eigenvector (recall that $s=e^{i{2\pi \/N}}$);
 $\cP_ku=\wt e_k(u,\wt e_k), u=(u_n)_1^N\in \C^N$ is a projector.
Define the operators $\wt \cS\p=(\cS \p_n)_{n\in \Z}$ and
 $\wt \cP_k\p=(\cP_k \p_n)_{n\in \Z}$.
 The operators $\wt \cS$ and $\mJ^{b}$ commute, then $\mJ^{b}=\os_1^N
(\mJ^{b} \wt \cP_k)$. Using \er{003arm}, \er{0031arm}  we deduce
that $\mJ^{b} \wt\cP_k$ is unitarily equivalent to the operator $
J_k^{b}$.\BBox

Below we use notation $a\ev a(b,v)$ and $d_n=d_n(b,v)$.

{\no\bf 2. The spectrum of unperturbed operator $ H^0$.}

We consider the case when all $v_{2n+1}=-v_{2n}=\wt v$ and $b=0$,
i.e. all $J_k$ are 1-periodic Jacobi matrices. For this case we
denote $\wt a=a(0,v)$, $\wt d=d_n(0,v)$. The monodromy matrix for $
J_k$ is
$$
  M_k(z)=\ma 0 & I_2\\ -(\wt a)^2 & \wt a(z-\wt d)
 \am=\left(\begin{array}{cccc}
 0   &   0   &   1   &   0  \\
 0   &   0   &   0   &   1  \\
 -s^k&   0   &-s^k   &(z-\wt v)s^k  \\
 0   &-s^k   &z+\wt v      & -1
 \end{array}\right).
$$
The determinant is
$$
  D_k(z,\t)=\det( M_k(z)-\t I_4)=\t^4+\t^3(s^k+1)+\t^2s^k(3+\wt v^2-z^2)+\t
 s^k(s^k+1)+s^{2k}
$$
$$
 =s^{2k}\lt(\wt\t^4+\wt\t^32c_k+\wt\t^2(3+\wt v^2-z^2)+\wt\t2c_k+1\rt)
$$
$$
 =s^{2k}\wt\t^2
 \lt((\wt\t+\wt\t^{-1})^2+2c_k(\wt\t+\wt\t^{-1})+1+\wt v^2-z^2\rt)=
$$
$$
 =s^{2k}\wt\t^2(\wt\t+\wt\t^{-1}-\D_k^-(z))(\wt\t+\wt\t^{-1}-\D_k^+(z)),\ \ \
 where\ \ \wt\t=s^{-\frac k2} {\t}\ \ and
$$
\[\lb{Lparm}
 \D_k^{\pm}(z)=\pm\sqrt{z^2-\wt v^2-s_k^2}-c_k,\ \ where\ \
 c_k=\cos\frac{\pi k}N,\ s_k=\sin\frac{\pi k}N.
\]
The spectrum of $ J_k$ is
\[ \lb{Jkz}
 \s( J^0_k)=\{z\in\R:\  D_k(z,\t)=0\ for\ some\ \t\in\S^1\}=
\]
$$
 =\{z\in\R:\
 -2\le\D_k^{\pm}(z)\le2\}=(-\s_k^1)\cup(-\s_k^2)\cup(\s_k^2)\cup(\s_k^1),\
 \ \ where
$$
$$
 \s_k^1=[\sqrt{\wt v^2+s_k^2},\sqrt{5+\wt v^2+4c_k}],\ \ \s_k^2=[\sqrt{\wt v^2+s_k^2},\sqrt{5+\wt v^2-4c_k}].
$$
The spectrum of $H$ is
\[\lb{Hkz00}
 \s(H)=\bigcup_{k=1}^N\s(J_k)=[-\sqrt{9+\wt v^2},\sqrt{9+\wt v^2}]\sm(-|\wt v|,|\wt v|).
\]
In particular case, if $\wt v=0$, then $\s(H^0)=[-3,3]$.

{\no\bf 3. Small $2p$-periodic real potentials.} We consider the
case $b=0$. Firstly let $J\ev J(q):\ell^2(\Z)\to\ell^2(\Z)$ is a
$p$-periodic Shr\"odinger operator, i.e
$$
 (J f)_n=f_{n-1}+f_{n+1}+q_nf_n,\ \ f=(f_n)_{n\in\Z}\in\ell^2(\Z),
$$
where $q=(q_n)_{n=1}^p\in\ell_{\R}^{\iy}(\Z)$ and $q_{n+p}=q_n$ for
all $n\in\Z$. It is well known (see \cite{KKu1}), that the spectrum
of this operator is absolutely continuous and has a form
\[
\lb{sarm1} \s(J)=\s_{ac}(J)=\cup_1^{p} \s_{n},\qq
\s_n=[z_{n-1}^+,z_{n}^-],\qq   n\in\N_{p},
\]
\[\lb{sarm2}
z_{0}^{+}<z_{1}^{-}\le z_{1}^{+}<z_{2}^{-}\le
z_{2}^{+}<...<z_{p}^{-}.
\]
We denote $z_n^{\pm}(q)\ev z_n^{\pm}$. Also we introduce spectral
gaps $\g_n\ev\g_n(q)$ as
\[\lb{shgap}
 \g_n=(z_n^-,z_n^+),\qq n\in\N_{p-1}.
\]
If $q=0$ then
\[\lb{shz}
 z_n^{\pm}(0)=-2\cos\frac{\pi n}p,\ \ n\in\N_{p-1},\ \
 -z_0^+(0)=z_p^-(0)=2.
\]
For sufficiently small $q$ we have (see \cite{KKu1})
\[\lb{shgapz}
 z_n^{\pm}(q)=-2\cos\frac{\pi n}p+\hat q_0\pm|\hat q_n|+O(\|q\|^2),\ \
 q\to0,\ \ n\in\N_{p-1},
\]
\[\lb{shgapz0}
 z_0^{+}(q)=-2+\hat q_0+O(\|q\|^2),\ \ \
 z_p^{-}(q)=2+\hat q_0+O(\|q\|^2),\ \
 q\to0,
\]
where we denote $\hat q_n=(q,\hat e_{n})$, $\hat
e_n=p^{-1}(\t_n^{2j})_{j=0}^{p-1}$, $\t_n=e^{\frac{i\pi n}p}$.

Introduce the set $\X_p\ss\R^p$ by
\[\lb{shset}
 \X_p=\{\sum_{n\le\frac p2}\a_n(\hat e_n+\hat e_{p-n}),\ \ all\
 \a_n\ne0\}.
\]

Now we compare the spectrum of $H^0_{zi}(v)$ (zigzag) and $
H^0_{ar}(v)$ (armchair).

\begin{theorem}\lb{Tcomp}
i) Let $v_{2n}=v_{2n+1}$, $v_{n+2p}=v_n$ for all $n\in\Z$. Let
$v^{ev}=(v_{2n})_1^p$ and $J\ev J(v^{ev})$, then
\[\lb{armshcomp}
 (\s(J)+1)\cup(\s(J)-1)\ss\s( H_{ar}^0).
\]
ii) Let $N\in3\Z$ and $v_{n+p}=v_n$ for all $n\in\Z$. Let
$v=(v_n)_1^p$ and $J\ev J(v)$, then
\[\lb{zigshcomp}
 \s(J)\ss\s(H_{zi}^0).
\]
\end{theorem}

{\bf Proof of Theorem \ref{Tcomp}.} i) In our case (see \er{Jkarm})
we have
$$
 a^{N}(0,v)=\ma 0 & 1\\ 1& 0 \am=C\ma -1 & 0 \\ 0 & 1 \am C^*,
$$
$$
 d_n(0,v)=\ma v_{2n} & 1 \\ 1 & v_{2n} \am=C\ma v_{2n}-1 & 0 \\ 0 & v_{2n}+1 \am C^*
$$
for some unitary matrix $C$ ($CC^*=I_2$). Then $ J^0_N$ (see
\er{Jkarm}) unitarily equivalent to $(J(v^{ev})-I)\os(J(v^{ev})+I)$,
where $I$ is identity operator on $\ell^2(\Z)$. The statement ii)
was proved in Theorem \ref{T3} (see also Remark 2) on page 5). \BBox

For example we describe the spectrum of $ H^0(v)$ (armchair) near
$z=0$ and near $z=\pm3$ for small potentials $v$ (recall that $\s(
H^0(0))=[-3,3]$).

\begin{theorem} \lb{armspec}
Let $v_{2n}=v_{2n+1}$, $v_{n+2p}=v_n$ for all $n\in\Z$ and denote
$v^{ev}=(v_{2n})_1^p$. Let $p>2N>4$ and
$r_{\pm}=2\cos\lt(\frac{\pi}3\mp\frac1{2N}\mp\frac1{6p}\rt)-1$. Then
for sufficiently small $v$ we have
\[\lb{arms0}
 \s( H^0)\cap[r_-,r_+]=\lt([r_-,r_+]\sm
 \bigcup_{|n-\frac p3|\le\frac{p}{2N}}(\g_n+1)\rt)\cup
 \lt([r_-,r_+]\sm\bigcup_{|n-\frac {2p}3|\le\frac{p}{2N}}(\g_n-1)\rt),
\]
where first set and second set in the union has multiplicity $2$.
Also let
$$
 \wt r_-=1+2\cos\lt(\frac{\pi}{2N}+\frac1{6p}\rt),\ \ \ \wt
 r_+=1+2\cos\frac1{6p}.
$$
Then for sufficiently small $v$ we have
\[\lb{arms3z}
 \s( H^0)\cap[-\wt r_+,-\wt r_-]=[-\wt r_+,-\wt r_-]\sm
 \bigcup_{1\le n\le\frac p{2N}}(\g_n-1),
\]
\[\lb{arms3zaa}
 \s( H^0)\cap[\wt r_-,\wt r_+]=[\wt r_-,\wt r_+]\sm
 \bigcup_{p-\frac p{2N}\le n\le p-1}(\g_n+1),
\]
where set on the right side has multiplicity $2$. Moreover if
$v^{ev}\in\X_p$ then all $|\g_n|\ne0$ in \er{arms0}-\er{arms3zaa}.
\end{theorem}

{\bf Proof of Theorem \ref{armspec}.} We consider only the statement
\er{arms3z}, the proof of other statements is similar. We have (see
\er{shz})
$$
 -3>-\wt r_+>z_1^-(0)-1>z_{[\frac p{2N}]}^+(0)-1>-\wt r_->
 z_{[\frac p{2N}]+1}^+(0)-1>(5+4c_1)^{\frac12}>-1.
$$
This inequalities shows (see \er{armshcomp}) that for sufficiently
small $v$ we have
\[\lb{armset00}
 \s( H^0)\cap[-\wt r_+,-\wt r_-]=\s( J^0_N)\cap[-\wt r_+,-\wt
r_-]=(\s(J)-1)\cap[-\wt r_+,-\wt r_-],
\]
since $[-\wt r_+,-\wt r_-]\cap J_k^0=\es$, $k\in\N_{N-1}$ (see
\er{Jkarm}, \er{Jkz}). Using identities \er{armset00} and
\er{sarm1}-\er{shgapz} we obtain \er{arms3z}. \BBox

Let $v$ be sufficiently small. We denote by $ G_{ar}$, $G_{zi}$ is a
maximal possible number of the open gaps on the edge of spectrum,
i.e. in the set $\s( H_{ar}^0)\cap[-3,-3+\a]$ and $\s(
H_{zi}^0)\cap[-3,-3+\a]$ respectively, where $\a$ is a some
sufficiently small value. Now we estimate $G_{ar}$, $G_{zi}$ for
sufficiently large period $2p$.

\begin{corollary}\it
 For sufficiently large $p$ we have
 $$
   G_{ar} = \frac p\pi
  \arccos\lt(1-\frac{\a}{2}\rt)+o(p),
 $$
 $$
  G_{zi}= \frac p\pi
  \arccos\lt(1-\frac{6\a-\a^2}{4}\rt)+o(p)
 $$
as $p\to\iy$.
\end{corollary}

{\no\bf 4. Large $4p$-periodic real potentials.} Now we consider
Shr\"odinger operator $H$ on armchair nanotube with large periodic
potentials. We show that in this case the structure of the spectrum
 is the same in the essential as for zigzag nanotube (see Theorem
\ref{T44}), but the Proofs are different.

\begin{theorem} \lb{Larmp} i) Let $v=(v_n)_{-\iy}^{+\iy}$ be a $4p$-periodic
($p>2$) real potential such that $v_i\ne v_j$, $1\le i\ne j\le4p$.
Let $\s(t)=\s( H^{b}(tv))$ and $\s_k(t)=\s( J_k(tv))$. Then
\[\lb{T4arm}
 \s(t)=\bigcup_{k=1}^N\s_k(t),\ \
 \s_k(t)=\bigcup_{j=1}^{4p}\s_{k,j}(t),
\]
where intervals $\s_{k,j}(t)$ satisfy
\[\lb{T4arm1}
 |\s_{k,j}|=\frac{4}{t^{2p-1}\prod_{n\in (\cQ_i\sm
 j)}(v_j-v_n)}+O(t^{-2p}),\ \ |\s_{k,j}-\wt\l_j|=O(t^{-3}),\ \
 t\to\iy,\ \ j\in\cQ_i.
\]
Here $\wt\l_j$ are defined in \er{L018} and $\cQ_{i}$ are defined in
\er{L005}.

Moreover, if $v_1^1<v_2^1<..<v_{4p}^1$ and $b$ is sufficiently
small, then all intervals $\s_{k,j}(t)$ are disjoint for
sufficiently large $t$.
\end{theorem}

{\bf Proof of Theorem \ref{Larmp}.} Recall that
\[\lb{L001}
 a\ev a_k=\ma 0 & e^{ib_1}s^k\\e^{-ib_2} &0 \am,\ \ s=e^{\frac{2\pi
  i}N},  \qq d_n\ev \ma v_{2n} & e^{ib_3}\\e^{-ib_3}&v_{2n+1} \am,\qq
 n\in\Z,\ \ aa^*=I_4.
\]
Also we use notation $d_n\ev d_n(v)$, where $v=(v_1,..,v_{4p})$. The
monodromy matrix for operator $ J_k(v)$ is
\[\lb{L002}
 M_k\ev M_k(z)\ev M_k(z,v)=\cM_{2p}..\cM_{1},\ \ \
 \cM_n=\ma 0 & I_2\\ -a_k^2 & a_k(z-d_n) \am.
\]
It is well known that
\[\lb{Ls}
 \s( J_k^{b})=\{z:\ \det(M_k(z)-\t)=0\ for\
some\ \t\in \S^{1}\}.
\]
Using \er{L002} we obtain
\[\lb{L003}
 M_k=\ma 0 & 0\\ 0 & a_{2p}(z-d_{2p})..a_1(z-d_{1})
 \am+\ma P_1 & P_2\\ P_3 & P_4 \am,
\]
where $P_j\ev P_j(z-d_{2p},..,z-d_1)$ is a $2\ts2$ matrix polynomial
and $\deg P_j < 2p$ for all $j=1,..,4$. Also, using \er{L001} and
periodicity of $v$, we deduce that
\[\lb{L004}
 a_{2p}(z-d_{2p})..a_1(z-d_{1})=(\det a_k)^p\ma \prod_{n\in \cQ_1}(z-v_n) & 0\\
 0 & \prod_{n\in \cQ_2}(z-v_n) \am+\ma Q_1 & Q_2 \\ Q_3 & Q_4 \am,
\]
where $Q_j\ev Q_j(z-v_{4p},..,z-v_1)$ are polynomials and $\deg Q_j
< 2p$, sets $\cQ_j$ are
\[\lb{L005}
 \cQ_1=\cup_{j=0}^{p-1}\{4j+1,4j+2\},\ \
 \cQ_2=\N_{4p}\sm\cQ_1=\cup_{j=0}^{p-1}\{4j+3,4j+4\}.
\]
Let $D_k(z,\t)\ev D_k(z,\t,v)=\det(M_k-\t I_2)$. Using
\er{L003}-\er{L005} we get
\[\lb{L006}
 D_k(z,\t)=\t^4+\det a_k^p\t^3(\prod_{n\in \cQ_1}(z-v_n)+\prod_{n\in
 \cQ_2}(z-v_n)+R_1)+\det a_k^{2p}\t^2(\prod_{n=1}^{4p}(z-v_n)+R_2)
\]
$$
 +\t\wt R_1+\wt R_2,
$$
where polynomials
\[\lb{L007}
 R_1\ev R_1(z-v_{4p},..,z-v_1),\ \ \deg R_1<2p,\ \
 R_2\ev R_2(z-v_{4p},..,z-v_1),\ \ \deg R_2<4p,
\]
\[\lb{L008}
 \wt R_1\ev \wt R_1(z-v_{4p},..,z-v_1),\ \ \wt
 R_2\ev R_2(z-v_{4p},..,z-v_1).
\]
are not depended on $\t$. Let $\t\in S^1$, $z\in\R$, then it is well
known, that the polynomial
\[\lb{L009}
 \wt D_k(z)\ev\wt D_k(z,\t)\ev\wt D_k(z,\t,v)=
 (\det a_k^{-2p})\t^{-2}D_k(z,\t)=\prod_{n=1}^{4p}(z-v_n)+O(z^{4p-1}),\ \
 z\to\iy.
\]
is real, since it has only real zeroes, because the spectrum of $
J_k$ is real. Let $\t\in \S^1$, $z\in\R$, then using \er{L009},
\er{L006} and $\overline{\wt D_k(z,\t)}\ev \wt D_k(z,\t)$,
$a_k^*=a_k^{-1}$ we deduce that
\[\lb{L010}
 \wt R_1=\det a_k^{3p}(\prod_{n\in \cQ_1}(z-v_n)+\prod_{n\in
 \cQ_2}(z-v_n)+\ol{R_1}),\ \ \wt R_2=\det a_k^{4p}.
\]
Substituting \er{L010} into \er{L006} and using \er{L009} we deduce
that
\[\lb{L011}
 \wt D_k(z)=\prod_{n=1}^{4p}(z-v_n)+R_2+2\Re(\t\det a_k^{-p})
 (\prod_{n\in \cQ_1}(z-v_n)+\prod_{n\in \cQ_2}(z-v_n))
\]
$$
 +2\Re(\t\det a_k^{-p}R_1)+2\Re(\t^2\det a_k^{-2p}).
$$
Now we denote $a=\frac1t$, $\l=\frac zt$ and $F_k(\l)\ev
F_k(\l,a)\ev F_k(\l,\t,a)=t^{-4p}\wt D_k(z,\t,tv)$. Then, using
\er{L011}, \er{L007}, we deduce that
\[\lb{L012}
 F_k=\prod_{n=1}^{4p}(\l-v_n)+aG_1(\l,a)+a^{2p}2\Re(\t\det a_k^{-p})
 (\prod_{n\in \cQ_1}(z-v_n)+\prod_{n\in \cQ_2}(z-v_n))
 +a^{2p+1}G_2(\l,a),
\]
where $G_1,G_2$ are polynomials and $G_1$ is not depended on $\t$.
Let $\l_j(a)\ev \l_j(a,\t)$ be zeroes of $F_k(\l)$ such that
$\l_j(0)=v_j^1$, these are analytic functions. Using similar
arguments as in "zigzag case", we deduce that derivatives
$(\l_j)^{(r)}(0)$ are not depended on $\t$ for all $j\in\N_{4p}$,
$r\in\N_{2p-1}$ and
\[\lb{L013}
 (\l_j)^{(2p)}(0)=\frac{-2\Re(\t\det a_k^{-p})}{\prod_{n\in (\cQ_i\sm
 j)}(v_j-v_n)},\ \ where\ j\in\cQ_i\ for\ some\ i=1,2.
\]
These yield
\[\lb{L014}
 |\l_j(a,\S^1)|=\frac{4a^{2p}}{\prod_{n\in (\cQ_i\sm
 j)}(v_j-v_n)}+O(a^{p+1}),\ \ a\to0,
\]
where $j\in\cQ_i$ for some $i=1,2$. Let $z_j(t)\ev z_j(t,\t)$,
$j\in\N_{4p}$ be zeroes of $D_k(z,\t,tv)$, then $z_j=t\l_j$ and
\[\lb{L015}
 |\s_{k,j}(t)|=|z_j(t,\S^1)|=\frac{4}{t^{2p-1}\prod_{n\in (\cQ_i\sm
 j)}(v_j-v_n)}+O(t^{-2p}),\ \ t\to\iy,
\]
where the spectrum $\s( J_k(tv))=\cup_1^{4p}\s_{k,j}(t)$. Introduce
the $\C^{4p\ts4p}$ matrices $L_{k}(\t)\ev L_k(\t,t)$ and
$B_{k}(\t)\ev B_k(\t,t)$ by
\[\lb{L016}
  L_k=B_k+\diag(tv)=\left(\begin{array}{ccccc} d & a_k^*& 0 & ... & {a_k\/\t} \\
                                   a_k & d & a_k^* & ... & 0 \\
                                  0 &  a_k & d & ... & 0 \\
                                  ... & ... & ... & ... & ... \\
                                  \t a_k^* & 0 & ... & a_k & d
                                  \end{array}\right)+\diag(tv),
\]
where
\[\lb{L017}
 d=\ma 0 & e^{ib_3}\\e^{-ib_3}&0 \am.
\]
Let $\wt\l_j(t)\ev\wt\l_j(t,\t)$ be eigenvalues of $L_k$, it is well
known, that $\s_{k,j}(t)=\l_j(t,\S^1)$. Then perturbation theory
gives us
\[\lb{L018}
 \wt\l_j=v_jt+(B_ke_j,e_j)-\frac1t\sum_{n\in\N_{4p}\sm
 j}(v_n-v_j)|(B_ke_j,e_n)|^2+...=
\]
$$
 =\ca v_jt-\frac{V_{j,-1}+V_{j,1}+V_{j,3}}t-
 \frac{\wt s_k(V_{j,-2}V_{j,-1}V_{j,1}+V_{j,1}V_{j,2}V_{j,3})}{t^2}+O(t^{-3}) & j\in2\N \\
      v_j^1t-\frac{V_{j,-3}+V_{j,-1}+V_{j,1}}t-
 \frac{\wt s_k(V_{j,-3}V_{j,-2}V_{j,-1}+V_{j,-1}V_{j,1}V_{j,2})}{t^2}+O(t^{-3}) & j\in2\N+1  \ac,
 \ \ t\to\iy,
$$
$$
 where\ \ V_{j,k}=(v_{j+k}-v_j)^{-1},\ \ \wt s_k=2\Re(s^k
 e^{i(b_1+b_2-2b_3)}).
$$

\BBox

\section{Appendix}
\setcounter{equation}{0}

Below we consider the unperturbed Jacobi operator $J^0(a)=J(a,0)$
 given by (see \er{j0001})
\[
\lb{j0001a}
 (J^0(a)y)_n=a_{n-1}y_{n-1}+a_ny_{n+1},\ \ a_{2n}=a>0,\ \
 a_{2n+1}=1,\ \ n\in\Z, \
 y=(y_n)_{n\in\Z}.
\]

\begin{lemma}
\lb{b001l} The eigenvalues $z_n^s$ and the eigenvectors $e_n^s,
(n,s)\in \N_{p}\ts \N_2$ of the matrix $K^0(e^{i\f},a)$, $\f\in\R$ (
given by \er{dL}) have the forms:

if $\ve_n=a+e^{ir_n}\ne 0, r_n={\f+2\pi n\/p}$, then
\[ \lb{t001}
 z_n^s=(-1)^s|\ve_n|,\ \
 e_n^s=(2p)^{-\frac12}(e_{n,j}^s)_{j=1}^{2p}\in \C^{2p},\\
 e^s_{n,2j}=(-1)^se^{ijr_n},\ \ e_{n,2j+1}=e^{ijr_n}\frac{\ve_n}{|\ve_n|}.
\]
If $\ve_n=0$, then the eigenvalue $z_n^1=z_n^2=0$ has the
multiplicity two and the corresponding orthogonal eigenvectors are
given by
$$
 e_n^1=(1,1,-1,-1,1,1,..)^{\top},\ \ e_n^2=(1,-1,-1,1,1,-1,..)^{\top}
 \in \C^{2p}.
$$
\end{lemma}

\no {\bf Proof.} We need the simple fact. Let $K^0(\t)e=z e$ for
some $z,\t$ and the eigenvector $e=(f_n)_1^{2p}$. Introduce two
numbers $
 f_0=\t^{-1}f_{2p},\ \ f_{2p+1}=\t f_1$.
Then
$$
 M_2(z)(f_{n-1},f_n)^{\top}=(f_{n+1},f_{n+2})^{\top},\ \
 M_2(z)^p(f_0,f_1)^{\top}=(f_{2p},f_{2p+1})^{\top}=\t(f_0,f_1)^{\top},
$$
and $(f_0,f_1)^{\top}$ is the eigenvector of the monodromy matrix
$M_2$ given by \er{M2} at $v=0$.

Conversely, let $M_2(z_1)(f_0,f_1)^{\top}=\t (f_0,f_1)^{\top}$ for
some $\t, z_1$. We introduce the vectors $
(f_{n+1},f_{n+2})^{\top}=M_2(z_1)(f_{n-1},f_{n})^{\top}, n\in
\N_{2p-2}. $ Then
\[
\lb{mL} K^0(\t,a)e_1=z_1 e_1,\qqq where \qq e_1=(f_n)_1^{2p}.
\]
Recall that (see \er{D2}) the Lyapunov function $\D_2$
(corresponding to $M_2$) is given by $\D={1\/2}\Tr
M_2(z)={1\/2a}(z^2-a^2-1)$. Using these arguments we will determine
the eigenvalues and the eigenvectors of the matrix $K^0(\t,a)$.
Firstly, let $z_s=z_s(r)$ be solutions of the equation $\D(z)=\cos
r$ for fixed $r\in \R$. Then $(z_s)^2=a^2+2a\cos r+1=|\ve|^2,\ \
\ve=a+e^{ir}$, which yields
$$
 z_1=-|\ve|,\ \ z_2=|\ve|.
$$
We will determine the eigenvectors of the monodromy matrix
 $M_2(z_s), s=1,2$ for the eigenvalue $\t=e^{ir}$, since $\D(z_s)=\cos r$.
Firstly, if $\ve\ne 0$, then we obtain
$$
 M_2(z_s)-e^{ir}I_2=\ma -a-e^{ir} & z_s \\
 -z_s & a+2\cos r-e^{ir} \am=\ma -\ve & (-1)^s|\ve| \\
 (-1)^{s+1}|\ve| & \ve \am,
$$
and the corresponding eigenvectors are given by
\[
 \lb{b010}
 \e^s=\left(\begin{array}{c}\e_1^s \\
 \e_2^s\end{array}\right)=\left(\begin{array}{c}(-1)^s \\
 \frac{\ve}{|\ve|}\end{array}\right),\ \ s=1,2.
\]
Define the vectors $e^s=(e_n^s)_1^{2p}$ by
\[
 \lb{b011}
 \ma e^s_{0}\\ e^s_{1}\am =\e^s(r),\qqq
 \ma e^s_{2j}\\ e^s_{2j+1}\am=
 M_2^j(z_s)\ma e^s_{0}\\ e^s_{1}\am=e^{ijr} \e^s(r).
\]
Then using \er{mL} we deduce that $K^0(e^{ipr},a)e^s=z_{s}e^s$,
where identities \er{b010}, \er{b011} give the components of $e^s$
by
$$
 e^s=(e^s_j)_{j=1}^{2p},\ \
 e^s_{2j}=(-1)^se^{ijr},\ \
 e_{2j-1}^{s}=e^{ijr}{\ve\/|\ve|},
$$
which yields \er{t001}, since solutions of the equation
$e^{ipr}=e^{i\f}$ has the form $r_n=\frac{\f}p+\frac{2\pi n}p$,
$n\in \N_p$.

Secondly, if $\ve=a+e^{ir}=0$, then we deduce that $a=1, e^{ir}=-1,
z_s=0$, $s=1,2$ and the matrix $M_2(z_s)-e^{ir}I_2=0$. The
corresponding eigenvectors have the forms $\e^1=(-1,1)^{\top}$,
$\e^2=(1,1)^{\top}$ and using arguments as above, we obtain the
proof of the case $\ve=0$. \BBox

\begin{corollary}
\lb{b012} \it The spectrum of the operator $J^0(a)$ given by
\er{j0001a} has the form
$$
\s(J^0(a))=\cup_{n=1}^{2p}\s^0_n,\ \ \s^0_n=[\l_{n-1}^+,\l_n^-],\ \
\l_n^{\pm}\ev\l_n^{\pm}(a)=z_n^{\pm}(a,0),\ \ \l_{2p}^-=-\l_0^+=a+1,
$$
$$
 \l_n^{\pm}=\n_n^{\pm}|a+e^{i\frac{\pi n}p}|,\ \
 \n_n^{\pm}=(\pm1)^{\d_{n,p}}\sign(n-p),\
 \ n\in \N_{2p-1},\ \ \sign(0)=1,
$$
where  $\l_{2n}^{\pm}$ (and $\l_{2n+1}^{\pm}$) are all eigenvalue of
the matrix $K^0(1,a)$ (and $K^0(-1,a)$) given by \er{dL}.
Corresponding eigenvectors of $K^0(1,a)$ (and $K^0(-1,a)$) are given
by
\begin{multline} \lb{j1112}
 Z_n^{\pm}\ev Z_n^{\pm}(a)={1\/(2p)^{1\/2}}(f_{j,n}^{\pm})_{j=1}^{2p},\ \
 f^{\pm}_{2j,n}=\n_n^{\pm}\t_n^{\pm j},\ \
 f_{2j+1,n}^{\pm}=\t_n^{\pm
 j}e^{\pm i\arg(a+\t_n)},\\
 \t_n=e^{\frac{i\pi n}{p}},\ \ j\in \N_p,\ \ a+\t_n\not=0,
\end{multline}
and
$$
 Z_n^+=(2p)^{-\frac12}(1,1,-1,-1,1,1,..)^{\top},\ \
 Z_n^-=(2p)^{-\frac12}(1,-1,-1,1,1,-1,..)^{\top},\ \ a+\t_n=0,
$$
and $\l_n^{-}(a)=\l_n^{+}(a)$, $n\in \N_{2p-1}\sm\{p\}$ has
multiplicity two. Also $\l_p^-(a)<\l_p^+(a)$, $a\ne1$ and
$\l_p^-(1)=\l_p^+(1)$. The vectors $Z_n^+$ and $Z_n^-,
n\in\N_{2p-1}$ are orthogonal.
\end{corollary}

\no {\bf Proof} follows from Lemma  \ref{b001l}. In particular, we
have the following identity $
 (2p)\langle Z_n^+,Z_n^-\rangle=(1+e^{2i\arg(a+\t_{n})})\sum_{j=1}^p\t_n^{2j}=0.
$ \BBox

{\bf 3D coordinates of $r_{\o}$ and $b_j$ in the case of armchair
nanotube.} We rewrite similar formulas from \cite{BK} adapted for
our case
\[\lb{3dco1}
 r_{n,j,k}=(R\cos\a_{n,j,k}, R\sin\a_{n,j,k}, nh),\ \ n\in\Z,\ \
 j\in\{0,1\},\ \ k\in\Z_N,
\]
where
$$
 \a_{2n,j,k}=\frac{2\pi(k-n)}N+\a_{0,j},\ \
 \a_{2n+1,j,k}=\frac{2\pi(k-n)}N+\a_{1,j},
$$
$$
 \a_{0,0}=2\wt\b,\ \ \a_{0,1}=\frac{2\pi}N,\ \ \a_{1,0}=\wt\b-\wt\a,\ \
 \a_{1,1}=\frac{\pi}N,
$$
$$
 \sin\wt\a=\frac1{2R},\ \ \sin\wt\b=\frac1R,\ \
 R=\frac{\sqrt{\cos\frac{\pi}N+\frac54}}{\sin\frac{\pi}N},
$$
$$
 h=\sqrt{2+R_1R_2-2R^2},\ \ R_{\wt j}=\sqrt{({\wt j}R)^2-1},\ \ \wt
 j=1,2,
$$
and the magnetic constants are
\[\lb{3dco2}
 b_1=b_2=\frac{B(R_2-R_1)}4,\ \ b_3=-\frac{BR_2}4.
\]

\end{document}